# Spectroscopy and Biosensing with Optically Resonant Dielectric Nanostructures

*Alex Krasnok\*, Martín Caldarola, Nicolas Bonod, and Andrea Alú\**

Dr. A. Krasnok, Prof. Dr. A. Alù

Department of Electrical and Computer Engineering, The University of Texas at Austin, Austin, Texas 78712, USA

E-mail: alu@mail.utexas.edu (A. A.), akrasnok@utexas.edu (A. K.)

Martín Caldarola

Huygens-Kamerlingh Onnes Laboratory, Leiden University, Leiden, Netherlands

Dr. Nicolas Bonod

Aix-Marseille Univ, CNRS, Centrale Marseille, Institut Fresnel, Marseille, France

**Abstract**

Resonant dielectric nanoparticles (RDNs) made of materials with large positive dielectric permittivity, such as Si, GaP, GaAs, have become a powerful platform for modern light science, enabling various fascinating applications in nanophotonics and quantum optics. In addition to light localization at the nanoscale, dielectric nanostructures provide electric and magnetic resonant responses throughout the visible and infrared spectrum, low dissipative losses and optical heating, low doping effect and absence of quenching, which are interesting for spectroscopy and biosensing applications. In this review, we present state-of-the-art applications of optically resonant high-index dielectric nanostructures as a multifunctional platform for light-matter interactions. Nanoscale control of quantum emitters and applications for enhanced spectroscopy including fluorescence spectroscopy, surface-enhanced Raman scattering (SERS), biosensing, and lab-on-a-chip technology are surveyed. We describe the theoretical background underlying these effects, overview realizations of specific resonant dielectric nanostructures and hybrid excitonic systems, and outlook the challenges in this field, which remain open to future research.

## 1. Introduction

The operation principles of plasmonic nanoantennas and nanostructures are based on the optical properties of metals (e.g., Au and Ag). They have a profound impact on nanophotonics, as they provide efficient means to manipulate light and enhance light-matter interactions at the nanoscale[1–10]. These nanostructures can significantly enhance the interactions between quantum emitters (e.g., quantum dots, defect centers in solid, molecules) and their surrounding photonic environment[11–14], leading to giant luminescence enhancement[9,15–24], ultrafast emission in the picosecond range[25–27], strong coupling[28,29], surface enhanced Raman scattering (SERS)[30–32], optical interconnections[33,34], and control of emission patterns[35–38]. Because of these advantages, plasmonic nanostructures are broadly used in many applications, including near-field microscopy[39–41], biosensors[42,43], photovoltaics[44,45], photodetection[46], and medicine[47,48]. However, the typical plasmonic materials, gold and silver, have finite conductivities at optical frequencies, leading to inherent dissipation of the electromagnetic energy. This dissipation is caused by interband transitions of valence electrons to the Fermi surface or the electrons near the Fermi surface to the next unoccupied states in the conduction band[49], which leads to Joule heating of the structure and its local environment. For many applications, heat generation in the nanostructure and its surroundings are detrimental, since the behavior of the molecules under study can be modified dramatically with temperature. Moreover, elevated temperatures may lead to reshaping and even complete destruction of the nanostructures. In addition, a quantum emitter placed in the nanometer proximity of a metallic nanostructure will be quenched due to the dominant non-radiative decay channels, leading to the necessity of various dielectric spacers that reduce the overall enhancement effects.

The issue of optical losses in plasmonic nanostructures has been addressed in many recent studies[49–55] and it is still an object of discussion. It has been shown that plasmonic resonators are always accompanied by losses because of their nature: in order to achieve the resonant behavior for subwavelength plasmonic resonators, part of the optical energy is stored in the kinetic energy of electrons[53]. Studies show that even high-$T_c$ superconductors cannot be a good alternative to noble metals because they would have to operate at energies less than the superconductive gap, that is, in the THz or far-IR[52]. Using highly doped semiconductors in place of metals is also not a panacea, because the Fermi level decline necessarily means a decrease of the plasma frequency, thus impacting surface and localized plasmonic resonances[53], at which noble metals demonstrate superior properties. We also note that there is a limit to the overall electric field enhancement in metal nanostructures because of Landau damping[53,54], which is also caused by high densities of free electrons.

To circumvent the issues of loss and heating, the use of high refractive index dielectrics such as Si, Ge or GaP has been proposed[56–68]. When compared to metallic nanostructures, the main difference resides in the significantly reduced absorption in these materials: while optical resonances of metallic nanoparticles (NPs) are based on conduction electron oscillations (plasmons), the optical properties of RDNs are based on polarization charges and thus present inherently lower Ohmic losses. Note that an equivalence between the electromagnetic fields yielded by plasmonic and Mie resonators was recently established theoretically in Ref.[69]. It has been demonstrated that NPs and nanostructures made of such dielectric materials provide low dissipative losses, resonant electric and magnetic optical response, moderate light localization at the nanoscale, and strong inherent Raman scattering, which are not typical for plasmonic counterparts[64,66,68,70]. Because of their distinct properties, RDNs have been proposed for high-harmonic generation[67,71–78], photonic topological insulators[79,80], boosting the luminescence from quantum emitters such as NV-centers in nanodiamonds[81], quantum dots[82–85], perovskites[86], dye molecules[87–90], and carbon nanotubes[91]. In addition, resonant dielectric nanostructures have been used for scattering engineering[92–98], ultrafast switchers and modulators[99–105], optical interconnections on a chip[106–109], light trapping structures[110–114], colored metasurfaces[115–118], enhanced Raman scattering[119–122]. Interesting photonic phenomena such as Fano resonances[123–127], Purcell effect[128–132], and strong coupling[133] have been shown in dielectric nanostructures as well.

In recent times, the vanguard of this highly developing research area has been shifting to spectroscopy and biosensing applications. It has been demonstrated that, unlike high-Q microcavities, RDNs have a broadband response and thus low-quality factors that are compensated by nanoscale field confinement that allows for high fluorescence enhancement. Moreover, the low absorption of these materials reduces quenching effects that can be detrimental for fluorescence enhancement applications. In this work, we review the state-of-the-art applications of optically resonant high-index dielectric nanoparticles as a multifunctional platform for light-matter interactions with a special emphasis on optical spectroscopy (**Figure 1**, left) and biosensing (**Figure 1**, right). We focus on nanoscale control of light emitters and its applications for enhanced spectroscopy, such as surface-enhanced Raman scattering (SERS), fluorescence spectroscopy, biosensing, including the state-of-the-art technology. In Section A., we summarize the main outcomes of Mie theory to describe light scattering on high-index RDNs. Then, we give a theoretical background in the areas of spontaneous emission, especially focusing on weak and strong coupling regimes, mode volume, and Purcell effect. Section B. is devoted to state-of-art studies on high-index nanoantennas and their application for quantum source light emission. Section C. surveys optical spectroscopy with RDNs, including surface-enhanced Raman scattering (SERS), many and single-molecule fluorescence. Section D is devoted to biosensors based on RDNs. Finally, we summarize these findings and present an outlook for the other effects like *in vivo* imaging, thermometry at the nanoscale, and single-molecule strong coupling achievable with high-index RDNs.

## A. Theoretical background

**Mie theory and nanoparticle heating**. The optical response of spherical metallic and dielectric NPs of arbitrary radius R can be calculated with the multipolar Mie theory developed by Gustav Mie in 1908[134]. Below, we give some general conclusions that follow from this theory. The field of an incident plane wave with electric field strength $\mathbf{E}_0$ scattered by a spherical particle made of a material with dielectric permittivity $\varepsilon = n^2$ ($n$ is the refractive index of the material) can be written as a sum of partial spherical waves with coefficients $a_l$ and $b_l$ that are electric and magnetic scattering amplitudes. The theory gives the following expressions for normalized scattering [ $Q_{sct} = P_{sct}/(\pi R^2 I)$ ], extinction [ $Q_{ext} = P_{ext}/(\pi R^2 I)$ ] and absorption [ $Q_{abs} = P_{abs}/(\pi R^2 I)$ ] cross sections for particles made of a nonmagnetic material[135–137]:

$$\begin{aligned}
Q_{sct} &= \frac{2}{(kR)^2} \sum_{l=1}^{\infty} (2l+1)\left(|a_l|^2 + |b_l|^2\right), \\
Q_{ext} &= \frac{2}{(kR)^2} \sum_{l=1}^{\infty} (2l+1)\operatorname{Re}(a_l + b_l), \\
Q_{abs} &= Q_{ext} - Q_{sct},
\end{aligned} \quad (1)$$

where $l$ defines the order of partial wave, $k$ is the wavenumber $\left(k = \frac{\omega}{c} n_h = \frac{2\pi}{\lambda} n_h\right)$, $\lambda$ is the wavelength in vacuum and $\varepsilon_h = n_h^2$ is the dielectric permittivity of the surrounding medium.

The quantities $P_{sct}$, $P_{ext}$ and $P_{abs}$ denote respectively the scattering, extinction and absorption power, $I$ is the excitation intensity ($I = n_h c \varepsilon_0 |\mathbf{E}_0|^2 /2$), $c$ is the speed of light, and $\varepsilon_0$ is the dielectric constant. Electric and magnetic scattering amplitudes are given by

$$a_l = \frac{R_l^{(a)}}{R_l^{(a)} + iT_l^{(a)}}, \quad b_l = \frac{R_l^{(b)}}{R_l^{(b)} + iT_l^{(b)}}, \quad (2)$$

where $R_l$ and $T_l$ can be expressed as

$$\begin{aligned}
R_l^{(a)} &= n\psi_l'(kR)\psi_l(nkR) - \psi_l'(nkR)\psi_l(kR), \\
T_l^{(a)} &= n\chi_l'(kR)\psi_l(nkR) - \psi_l'(nkR)\chi_l(kR), \\
R_l^{(b)} &= n\psi_l'(nkR)\psi_l(kR) - \psi_l'(kR)\psi_l(nkR), \\
T_l^{(b)} &= n\chi_l(kR)\psi_l'(nkR) - \psi_l(nkR)\chi_l'(kR),
\end{aligned} \quad (3)$$

Here, $\psi_l(x) = \sqrt{\frac{\pi x}{2}} J_{l+1/2}(x)$, $\chi_l(x) = \sqrt{\frac{\pi x}{2}} N_{l+1/2}(x)$, $J_{l+1/2}(x)$ and $N_{l+1/2}(x)$ are the Bessel and Neumann functions, and the prime means derivation[135].

The equations $R_l^{(a)} + iT_l^{(a)} = 0$ and $R_l^{(b)} + iT_l^{(b)} = 0$ define the complex resonant frequencies $\omega_l$ of the electric and magnetic eigenmodes of the particle, respectively. These eigenmodes manifest themselves at real frequencies as resonant enhancement of scattering and extinction cross section spectra. In the general case, these expressions are satisfied at complex frequencies $\omega_l = \text{Re}[\omega_l] + i\,\text{Im}[\omega_l]$. Because the energy leaks out of the particle and is dissipated into absorption, the modes possess a finite lifetime $\gamma_{cav} = 1/\text{Im}[\omega_l]$, usually known as the *cavity decay rate*, which determines the half-width of resonance line at half maximum[138]. The resonance is usually described by its *quality factor*, which is defined as $Q = \omega / 2\gamma_{cav}$. In general, the cavity decay rate includes both radiative and absorptive parts: $\gamma_{cav} = \gamma_{cav}^{rad} + \gamma_{cav}^{loss}$, such that the quality factor can be expressed as $Q^{-1} = Q_{rad}^{-1} + Q_{loss}^{-1}$ [139].

Using Eqs. (1)-(3), we perform a comparative analysis of the scattering and absorption cross sections of plasmonic and high-index dielectric NPs. For that purpose, we use silver (Ag) which has the lowest absorption levels in the optical range among other noble metals[140].

We assume that the RDN is made of crystalline silicon (c-Si), which also exhibits the best performance in the visible region[141]. The calculation results are shown in **Figure 2**. The Ag NP features only electric-type resonances, including electric dipole (ed), electric quadrupole (eq), and electric octupole (eo), whose dispersion with increasing radius (R) is schematically shown by white dashed lines. On the contrary, the spectrum of resonant modes of Si NP is richer and it includes the magnetic dipole (md) moment, as well as magnetic quadrupole (mq) and higher order resonant modes. We note that although more sophisticated plasmonic structures can support modes of magnetic nature[142,143], they suffer from dissipative losses and complexity in fabrication. **Figures 2**(b), (d) show the calculated normalized absorption cross sections spectra of (b) the Ag NP and (d) the Si NP. A comparative analysis of these results shows that the Si NP has significantly less dissipative losses in the entire range of parameters. In addition, the dissipative losses of the Si NP rapidly fall as the wavelength increases, which is explained by the lower losses of Si away from the main absorption band. Note that gold (Au) NPs have even greater dissipative losses than Ag, while the former are more stable in ambient conditions[144].

The emitted or scattered electromagnetic energy in a given direction is described by the *directivity* $D$[3,5,145], whose standard definition is given by

$$D(\theta,\varphi) = \frac{4\pi P(\theta,\varphi)}{P_{tot}}, \tag{4}$$

where $P_{tot} = \pi R^2 I Q_{sct}$ is the total radiated or scattered power into the far-field, i.e., the integral of the angular distribution of the radiated power over the spherical surface, $(\theta,\varphi)$ are the standard angular coordinates of the spherical coordinate system. The directivity quantifies the modification that must be experienced by the emitted power to obtain in the same direction the same Poynting vector modulus yielded by an isotropic source at the same point. This expression is normalized to $4\pi$, thus the directivity of an isotropic source equals unity, and the one of a small dipole is 1.5. The directivity is a function of the direction in space, but sometimes it is enough to know the maximum value of this quantity, namely, its value in the direction of a main lobe of power pattern: $D_{max} = 4\pi \max[P(\theta,\varphi)]/P_{tot}$.

Single NP scattering measurements are usually conducted by so-called *dark-field microscopy*, which consists in a selective collection of the scattered light, over a dark background, i.e., in directions not illuminated by the incident field[146]. Mathematically, it implies the measurement of the scattered light in some directions that are not coincident with the forward one. For this reason, it is useful to present the equation for normalized scattering cross section into any direction ($\theta$)

$$Q_{sct}(\theta) = \frac{2}{(kR)^2}[i_1(\theta) + i_2(\theta)],$$

$$i_1(\theta) = \left|\sum_{l=1}^{\infty}\frac{2l+1}{l(l+1)}[a_l\pi_l(\theta) + b_l\tau_l(\theta)]\right|^2, \quad (5)$$

$$i_2(\theta) = \left|\sum_{l=1}^{\infty}\frac{2l+1}{l(l+1)}[a_l\tau_l(\theta) + b_l\pi_l(\theta)]\right|^2,$$

where $\pi_l(\theta) = \frac{P_l^{(1)}(\cos(\theta))}{\sin(\theta)}$, $\tau_l(\theta) = \frac{dP_l^{(1)}(\cos(\theta))}{d\theta}$, and $P_l^{(1)}$ is the associated Legendre polynomial. Usually NPs are deposited on a dielectric substrate, providing them with an optically inhomogeneous environment. Such dielectric substrate can break the symmetry and, thus, a dependence of the scattered signal on ($\varphi$) may appear.

The absorption cross section $Q_{abs} = P_{abs}/(\pi R^2 I)$ defines the power absorbed by a particle made of a material with thermal conductivity $\chi_1$ located in a medium with thermal conductivity $\chi_2$ and, thus, its temperature increase $\Delta T$ in the steady-state regime[147]:

$$\Delta T = R\frac{Q_{abs}}{4\chi_2}I. \quad (6)$$

For a discussion of this expression and its scope of applicability, see Ref.[147]. Thus, dielectric nanostructures located in a medium with low thermal conductivity ($\chi_2$) may be heated, which has been already proven to be an effective way for the creation of tunable nanophotonic devices[148]. Moreover, it has been demonstrated that, when optical fields are concentrated into nanoscale volumes, the speed of thermo-optical effects can approach the picosecond scale paving the way to ultrafast and ultrastrong nonlinear effects[149].

**Figure 3** shows the temperature increase after heating of (a) Ag and (b) Si spherical NPs in air ($\chi_2 = 0.025$ W/m·K [150]), as a function of their radius R and wavelength. The light intensity is assumed constant $I = 0.1$ mW/μm$^2$. At wavelengths below 400 nm, both Ag and Si NPs exhibit temperature increase up to 250 K. However, the temperature increase for the magnetic dipole resonance in Si NP is significantly reduced as the wavelength increases. At wavelengths larger 700 nm, the heating of Si NP is vanishingly small, whereas the Ag NP loses its resonant properties. For experimentally measured temperature increase in plasmonic and dielectric nanoantennas see Section C.

**Figure of merit for sensors**. The dependence of the plasmonic and Mie resonance frequencies on the refractive index of a surrounding medium (like a solvent or human serum) are ideal for sensing applications. The figure of merit (FoM) of such sensors is commonly defined as the resonance shift (spectral sensitivity, S) upon a change in the refractive index of the surrounding dielectric normalized by the resonance line width ($\Delta\lambda$)[151–156]:

$$\text{FoM} = \frac{S}{\Delta\lambda}. \tag{7}$$

**Figures 2**(e), (f) show the normalized scattering spectra of the Ag NP with radius of 30 nm (e) and Si NP of radius 85 nm placed in air ($\varepsilon_h = 1$, blue curves) and water ($\varepsilon_h = 1.77$, red curves). These results demonstrate that the plasmonic resonance of the Ag NP exhibits a red shift when the dielectric permittivity of the host medium increases. The Si NP (as well as any RDNs) also exhibits a red shift, but the shifting in this case is much smaller. However, other geometries can be used to achieve the required sensitivity for real-world applications combined with low heat generation. The operation principle of RDN biosensors are based on the influence of the environment on collective electromagnetic modes of NP arrays. It has been demonstrated that they have sufficiently high values of FoM (the record value is 103), even exceeding the one of plasmonic-based biosensors (see Section D). High-index dielectric NPs have other advantages over plasmonic ones, including high thermal stability and a wider range of surface functionalization mechanisms. Dielectric materials can be easily functionalized for biosensing applications with the groups –$NH_2$, -$COO^-$, -$PO_4^{3-}$, -$SH$ and -$OH$, just to name a few (this issue has been addressed in Ref.[157]).

**Weak and strong coupling regime**. Scattering and luminescence-based optical spectroscopy, especially when dealing with a few or even single molecules, rely on the ability of nanoscale resonators (hereinafter referred to as nanocavities) and their arrays to enhance the scattered and luminescence intensity. Any quantum emitter (e.g., an atom, quantum dot, molecule, hereinafter referred to as molecule) being placed close to a nanocavity may experience its influence, which can be distinguished into *coherent* and *incoherent*[158].

Coherent effects include processes in which the light scattered by a molecule can interfere with light scattered by a resonant nanoparticle. As a widespread example of such a process, we describe the *Fano resonance*. This concept was first introduced by Ugo Fano[159], and since then the Fano resonance has been observed in a wide range of optical systems. Generally, a Fano resonance emerges from the interference between a discrete and a continuum state, and leads to an asymmetric spectral lineshape due to the constructive and destructive interference of radiation from these two states. The Fano resonance in photonic nanostructures has been extensively studied in various systems[160–164]. In nanocavity-molecule structures the Fano resonance is caused by the coupling between at least one broad plasmonic mode and one narrow excitonic resonance[6,165–168] It has been shown that the Fano resonance can induce extra narrow peaks and dips in the optical spectra and it has a great potential for biosensing due to its high sensitivity to environmental changes.

Incoherent effects include (i) altering the emission rate and its spectra, and (ii) different nonradiative decay processes through energy and/or charge transfer[169]. Below we focus on the first effect, while the interested reader may refer to Refs.[170–175] for the second effect. In addition, a plasmonic nanocavity may initiate different chemical reactions[176], which are also out of the scope of this Review.

Altering the excitation rate relies on the ability of a nanocavity to localize light at subwavelength scales, overcoming the diffraction and single molecule detection limits[177]. In other words, an optical nanocavity changes the electric and magnetic fields in its vicinity. Since electric dipole transitions are usually much stronger than magnetic ones in small optical emitters, we consider a molecule with an induced electric dipole moment. Below, we will also mention a special type of materials with forbidden electric dipole transitions. The electric field enhancement factor $|\mathbf{E}/\mathbf{E}_0|$, with $\mathbf{E}$ and $\mathbf{E}_0$ being the electric field strength near a nanocavity and the impinging one, defines the absorption cross-section of a molecule (*excitation rate*), which can be written out in form

$$\sigma = \sigma_0 \frac{|\mathbf{n}_d \cdot \mathbf{E}|^2}{|\mathbf{n}_d \cdot \mathbf{E}_0|^2}, \qquad (8)$$

where $\sigma_0$ is the molecule absorption cross-section in the absence of a nanocavity, $\mathbf{n}_d$ is the orientation of dipole (it is assumed that the nanocavity does not change the molecule structure). Thus, since nanocavities may enhance the near-field distribution in a certain space region, the absorption cross-section of a molecule placed in this region is also enlarged. This gives rise to molecule excitation and overall luminescence intensity enhancement.

From the perspective of enhancement by individual NPs, the absence of size cut-off frequency for dipole plasmon resonance of plasmonic NPs results in much stronger field enhancement in their vicinity as compared to dielectric ones with same sizes in the sub-100-nm range. However, larger Si NPs, possessing a magnetic Mie-type resonance at the optical frequencies yield comparable near-field enhancement[70,178]. This effect was proved in bulk SERS experiments, where resonant Si NPs produce larger SERS effect as compared to Au ones of the same sizes[179], see Section C.

The coupling Hamiltonian of a hybrid system of coupled nanocavity (plasmonic or dielectric) and exciton subsystem in the Jaynes-Cummings formalism[158] is defined by the matrix element

$$H_{ij} = \mathbf{d} \cdot \mathbf{E}_v(\mathbf{r}_0) = \hbar g, \quad (9)$$

where $\mathbf{d} = e\langle j|\hat{\mathbf{r}}|i\rangle$ is the transition dipole moment of a two-level exciton subsystem, $e$ is the elementary charge, $\hat{\mathbf{r}}$ is the radius-vector operator, and $\mathbf{E}_v(\mathbf{r}_0)$ is the *vacuum* electric field of the resonator mode at the molecule position, $\hbar$ is the reduced Planck constant. The value $g$ is called *coupling constant* and has a special importance for the theory of coupled nanophotonic systems.

There are two characteristic regimes of spontaneous emission of a molecule coupled to a dielectric or plasmonic nanocavity. Namely, if the coupling constant is less than the total *decay rate* $\gamma = \gamma_{cav} + \gamma_{mol}$ of a system, ($g < \gamma$), the system is in the so-called *weak-coupling regime*. In this regime, there is no coherent energy exchange between the molecule and nanocavity, and the variation of spontaneous emission can be described by the *Purcell factor*, which will be defined below. If the opposite condition is satisfied ($g > \gamma$), the system becomes *strongly coupled*, which leads to a splitting of the scattering spectrum, which is known as *Rabi splitting*[29,180]. In the strong-coupling regime, the system demonstrates energy exchange between the molecule and the resonator (Rabi oscillations), resulting in a more complex behavior of the spontaneous emission, which becomes non-exponential.

The strong-coupling regime is very attractive for quantum optics applications, including the realization of hybrid electron-photon quasiparticles[181], Bose-Einstein condensates[182,183], single-photon switches[184,185], all-optical logic[186,187], and control of chemical reactions[188,189]. Moreover, the strong coupling regime has a special interest in quantum optics and nanophotonics if the system can be tuned at time scales less than the characteristic time of energy transfer between the cavity mode and the molecule, which is described by the inverse coupling constant (1/g).

*How can the strong coupling be achieved in an arbitrary system?* Since the coupling constant is proportional to the vacuum electric field of the cavity mode as $g = \frac{d}{\hbar}\sqrt{\frac{\hbar\omega}{2V_{eff}\varepsilon_0\varepsilon}}$ [14,190], in order to increase $g$ one can increase the dipole moment of the exciton subsystem and/or reduce the effective mode volume $V_{eff}$, which can be defined in the low loss limit as[14,191]

$$V_{eff} = \frac{\int \varepsilon(r)|E(r)|^2 \, dV}{\max\left(\varepsilon(r)|E(r)|^2\right)}. \tag{10}$$

Usually, the effective mode volume is much less than the physical volume of the resonator. For example, plasmonic nanocavities demonstrate unprecedented effective mode shrinking, up to values in the order of $\lambda^3/10^4$ [138]. This equation (10) is not always applicable to nanocavities (especially to plasmonic ones) because they often demonstrate high dissipative rates[192]. This issue has been addressed many times in the literature[138,192–197]. In particular, a detailed analysis for relatively small plasmonic nanocavities without gain has been presented in Ref.[138], leading to the concept of complex mode volume.

In the scattering spectra, the strong coupling regime appears as an anticrossing at the frequency of excitonic resonance. To simulate the coherent scattering spectra of the system the *coupled harmonic oscillator model* can be applied[198–201]. This model considers a cavity driven by an external field, which is coupled to the molecule (exciton resonance). According to this theory, in the harmonic oscillator approximation the scattering cross section of the hybrid system with two coupled oscillators is given by

$$\sigma_{sct} \propto \left|\frac{\omega^2\tilde{\omega}_{mol}^2}{\tilde{\omega}_{cav}^2\tilde{\omega}_{mol}^2 - g^2\omega^2}\right|^2, \tag{11}$$

where $\tilde{\omega}_{mol}^2 = \omega^2 - \omega_{mol}^2 + i2\gamma_{mol}\omega$ and $\tilde{\omega}_{cav}^2 = \omega^2 - \omega_{cav}^2 + i2\gamma_{cav}\omega$ are the harmonic oscillator terms for the molecule and nanocavity (plasmonic or Mie) resonances, respectively, with $\omega_{mol}$, $\omega_{cav}$ and $\gamma_{mol}$, $\gamma_{cav}$ being the corresponding resonance frequencies and dissipation rates (half width at half maximum). One can use this equation to fit the experimentally measured data and get the coupling constant in a considered system.

In the weak-coupling regime, the hybrid system of coupled nanocavity and molecule experiences a modification of the spontaneous emission rate $\gamma_{mol}$ of the initially excited molecule without altering the resonant frequencies of the constituting elements. This effect can be quantified by the *Purcell factor*, defined in Eq. (12). Similar to the case of nanocavity, a molecule in free space usually has both dissipation channels, radiative and absorptive, leading to $\gamma_{mol}^0 = \gamma_{mol}^{0,rad} + \gamma_{mol}^{0,loss}$. Being brought into interaction with a nanocavity, the spontaneous emission rate changes according to the Fermi golden rule $\gamma_{mol} = \frac{2\pi}{\hbar^2}|H_{ij}|^2 D$, where $D = 1/(\pi\gamma_{cav})$ is the value of *density of states* on resonance assuming that the nanocavity mode has a Lorentzian profile of width $2\gamma_{cav}$. Using Eq. (9), this equation can be expressed as $\gamma_{mol} = 2g^2/\gamma_{cav}$, and normalizing it by the molecule decay rate in free space $\gamma_{mol}^0 = \frac{nd^2\omega^3}{3\pi\hbar\varepsilon_0 c^3}$, we define the Purcell factor ($F_p$), written in two forms, as a function of the coupling constant $g$ and of the quality factor $Q$[14,202]:

$$F_p \equiv \frac{\gamma_{mol}}{\gamma_{mol}^0} = \frac{2g^2}{\gamma_{cav}\gamma_{mol}^0} = \frac{3}{4\pi^2}\left(\frac{\lambda}{n}\right)^3 \frac{Q}{V_{eff}}. \qquad (12)$$

Thus, the Purcell effect results in a modification of the spontaneous emission rate $\gamma_{mol}$ of a molecule induced by its interaction with an inhomogeneous environment (nanocavity), and is quantitatively expressed by the Purcell factor, Eq. (12)[128,129,138,203–205]. This modification is significant if the environment is a resonator tuned to the molecule emission frequency. Open nanocavities such as plasmonic or RDN nanoantennas can change the spontaneous emission lifetime of a single quantum emitter, that is very useful in spectroscopy and microscopy of single NV centers in nanodiamonds[206], Eu$^{3+}$-doped nanocrystals[207], optical sensing[138,208], and bioimaging[17].

In general, the Purcell effect gives rise to enhancement of emission and dissipation rates. Taking into account that usually one is interested in the radiative part of spontaneous emission, it is useful to distinguish the radiative ($F_p^{rad} = \gamma_{mol}^{rad} / \gamma_{mol}^0$) and dissipative ($F_p^{loss} = \gamma_{mol}^{loss} / \gamma_{mol}^0$) parts of the Purcell factor: $F_p = F_p^{rad} + F_p^{loss}$. In order to characterize the fraction of energy emitted into photons one also defines the *quantum yield* (or *quantum efficiency*) defined as $\eta = \gamma_{mol}^{rad} / (\gamma_{mol}^{rad} + \gamma_{mol}^{loss} + \gamma_{mol}^{0, loss})$. By introducing the vacuum quantum yield $\eta_0$ ($\eta_0 \leq 1$), which reflects the fraction of energy emitted by an isolated molecule, the resulting quantum yield $\eta$ ($\eta \leq 1$) of an emitter in the environment can be expressed through the Purcell factor:

$$\eta = \eta_0 \frac{F_p^{rad}}{\eta_0 F_p + (1-\eta_0)}. \tag{13}$$

Therefore, the fluorescence enhancement factor $\eta_{fl}$ is jointly affected by both the excitation enhancement, and the emission modification can be expressed as

$$\eta_{fl} = \sigma F_p^{rad} \eta, \tag{14}$$

such that the measured luminescence intensity $I_{fl}$ equals $I_{fl} = \eta_{fl} \varepsilon_{coll} I$, where $I$ is the intensity of excitation light, and by introducing the *collection efficiency* $\varepsilon_{coll}$ we take into account that only a part of emitted photons can reach a photodetector.

Concluding this Section we note that different combinations of the quality factor ($Q$) and effective mode volume ($V_{eff}$) define not only the Purcell effect ($\propto Q/V_{eff}$) and strong coupling ($\propto 1/V_{eff}$) but also many other optical processes, including optical bistability and single-photon Kerr nonlinearity[209]. This makes resonant nanostructures with designable $Q$ and $V_{eff}$ vital for many applications.

## B. Resonant dielectric nanostructures and hybrid excitonic systems

**Resonant dielectric nanoantennas and nanostructures**. Dielectric particles made of titania were proposed in 2010 to design highly directive lens antennas[210]. The authors suggested to couple dielectric materials to yield high directivity and plasmonic particles to yield high decay rates. Low refractive index microspheres were proposed a few years earlier to probe fluorescent signals of molecules[211–213] and high resolution microscopy[214]. Then, in 2011 it has been shown that one Si NP can act as a *Huygens element* in the optical range (**Figure 4**(a))[215], demonstrating so-called *Kerker effect*[216–222]. It has been demonstrated that such nanoantennas are able to switch the radiation pattern between forward and backward directions due to the presence of electric and magnetic resonant modes (see Section A.). Then, in Ref.[223] the high directionality of *Yagi-Uda nanoantenna*[145] composed of Si NPs has been shown, **Figure 4**(b). The optimal performance of the Yagi-Uda nanoantenna has been achieved for directors (particles 2-5) radii corresponding to the magnetic resonance (70 nm), and the radius of reflector (particle 1) corresponding to an electric resonance (75 nm) at a given frequency. This solution ensures that the directors and reflector operate in opposite reactive regimes (inductive and capacitive), consistent with antenna theory[224–227]. In **Figure 4**(b) the directivity of all-dielectric Yagi-Uda nanoantennas as a function of wavelength for a separation distance between particles of 70 nm is presented. The insets demonstrate the 3D radiation patterns at particular wavelengths. A strong maximum of the nanoantenna directivity (see Eq. (4)) reaching 12 at the wavelength of 500 nm has been achieved. By comparing plasmonic and all-dielectric Yagi-Uda nanoantennas, it has been demonstrated that the all-dielectric ones may exhibit better radiation efficiency, also allowing more compact design. A comprehensive analysis of such nanoantennas has been presented in Ref.[228].

In Ref.[229] it has been shown that more complex core-shell structures consisting of Ag core and Si shell may be adjusted to satisfy the resonant Kerker condition, possessing both resonant and unidirectional scattering at the same wavelength. The directionality of the structure can be further improved by arranging the core-shell NPs in an array.

These ideas were developed in a series of subsequent works[128,228,230–235]. It has been demonstrated that the unique optical properties and low dissipative losses make RDNs perfect candidates for the design of high-performance nanoantennas, low-loss metamaterials, and other novel all-dielectric nanophotonic devices. The key to such novel functionalities of high-index dielectric nanophotonic elements is the ability of subwavelength dielectric NPs to support simultaneously both electric and magnetic resonances, which can be controlled by their shape[236,237]. Moreover, the electric and magnetic field localization in the vicinity of all-dielectric nanoantennas has been theoretically predicted[178] and experimentally measured in the microwave[238], terahertz[239] and visible[240] ranges that has paved the way to applications of such systems for spectroscopy and sensing.

The existence of magnetic modes in RDNs provides a very powerful tool for the creation of highly directive nanoscale light sources. In Refs.[92,241] the concept of so-called *superdirective nanoantennas* based on the excitation of higher-order magnetic modes in RDNs of radius 90 nm has been proposed. The superdirective regime (highly directive radiation by an antenna of subwavelength dimensions) is achieved by placing an emitter within a small notch created on the nanosphere surface. It turns out that such a small modification of the sphere allows for efficient excitation of higher-order modes. Then, it has been demonstrated that this effect can be applied for emission extraction enhancement and to simplify the initialization of a single emitter in an assemble with a subwavelength separation among individual emitters[81]. In Ref.[242] it has been demonstrated that mixing of dipolar and quadrupolar modes may also lead to highly directive emission to perform an efficient switch of the emission directivity with respect to the wavelength.

Mie resonances in dielectric particles can also increase the Purcell effect, associated with either electric or magnetic transition rates nearby quantum emitters. Their large quality factors compensate the low field confinement as compared to the plasmon resonances of metallic nanostructures, for which nonradiative decay channels dominate. In Ref.[230] it has been shown theoretically that near-infrared quadrupolar magnetic resonances in Si NPs can preferentially promote magnetic versus electric radiative exciton in trivalent erbium ions at 1.54 μm (see **Figure 4**(c,d)). The distance-dependent interaction between magnetic (electric) dipole emitters and induced magnetic or electric dipoles and quadrupoles has been derived analytically and compared to full-field calculations based on Mie theory. Single Si NP can give rise to emission enhancement of up to 90 times for longitudinal magnetic dipole emitters and up to 15 times for transverse electric emitters. For simulated Purcell factor enhancement of dielectric dimer nanoantennas and their comparison with plasmonic ones see **Section C.**, where the Purcell factor enhancement up to ~ 200 times using RDNs is reported. Further increase of the Purcell factor (up to several hundreds) can be achieved by relying on the *Van Hove singularity* of a chain of high-index nanoparticles[129]. Such approach allows to substantially enhance the local density of optical states at the location of a quantum source and thus achieve high values of the Purcell factor with relatively small dielectric nanostructures, while leaving unaltered all their other advantages.

Recently, it has been theoretically demonstrated that hybrid metal-dielectric nanoantennas based on dielectric particles on metallic films yield high Purcell factor (>5000), high quantum efficiency (>90%) and highly directional emission at visible and infrared wavelengths[132]. A similar design of highly efficient metal-dielectric patch antennas for single-photon emission has been proposed in Ref.[243]. Such *hybrid nanostructures* may benefit from the use of both approaches, plasmonic and dielectric, and currently are a subject of intense studies.

In order to get a strong local field enhancement, plasmonic dimers[15,17,18,21] or oligomers can be used[244–248]. The same approach is also possible for dielectrics,[125,240,249,250] when local field enhancement factor in the gap of a Si dimer may be more than one order of magnitude[240]. Such enhancement was applied to achieve high SERS and enhanced luminescence effects[70,87,89,178]. Thus, high-index RDNs also provide field enhancement and Purcell effects that are large enough for detection of a few or even single molecules[88] (see Section C).

As we already mentioned above, high-index dielectric NPs possess electric and magnetic resonant optical responses. Therefore, they provide a unique opportunity to enhance the photoluminescence intensity from naturally weak magnetic dipole transitions, giving an additional spectroscopy tool[130,251–255]. Moreover, it has been shown that there are materials with forbidden dipole transitions to the first order, e.g., some transition metals such as $Cr^{3+}$ and lanthanide series ions such as $Eu^{3+}$ and $Er^{3+}$. Thus, the combination of such materials with RDNs may pave the way to fundamentally novel optical phenomena. Dielectric structures supporting strong magnetic field enhancement $|\mathbf{H}/\mathbf{H}_0|$ factors (up to 20 and even more) have been proposed[130,256], with $|\mathbf{H}|$ and $|\mathbf{H}_0|$ being the magnetic field strength near a nanostructure and the one of the impinging wave.

Progress in dielectric nanoantenna research is accompanied by developments in their fabrication methods. **Figure 5** displays various realized dielectric nanoantennas and resonant nanostructures fabricated by different methods, including single Si spheres (**Figure 5**(a)) and Si dimers (**Figure 5**(b)), fabricated by femtosecond-laser ablation[60,125] and chemical means[125], single Ge disks (**Figure 5**(d)) and Si disk dimers (**Figure 5**(e)), fabricated by electron beam lithography[70,257]. Recently, disk dimers have been also realized in form of split GaAs nanoposts[258]. More complex RDNs, including all-dielectric oligomers (**Figure 5**(c))[259] and metasurfaces (**Figure 5**(f))[260], have also been produced by electron beam lithography. Moreover, different laser assisted methods have been also developed[60,261,262]. For a comprehensive review of available materials and fabrication techniques, readers can be addressed to Ref.[68].

**Hybrid dielectric nanocavity-excitonic systems**. Nanostructures, which provide resonant optical modes, are the first constituting part of hybrid nanocavity-excitonic systems. The second part is an excitonic system, which support resonant electron transitions between two or even more well-defined quantum levels. Here we summarize all realized different hybrid RDN-excitonic systems, including dye molecules, J-aggregates, perovskites and transition metal dichalcogenides, **Figure 6**.

**Figure 6**(a) shows PL enhancement factors in hybrid Si NP & *dye molecules* system as a function of corresponding scattering intensities obtained for nine single Si spheres[90]. It has been shown that Si NP may operate as nanoantennas for PL enhancement and a single Si NP can enhance dye PL at maximum 200-fold. The *strong coupling* regime with splitting energy of 100 meV has been demonstrated in hybrid Si NP & *J-aggregate* system[133], **Figure 6**(b). The black solid lines are fitting results according to the coupled harmonic oscillator model (see Eq. (11)).

It has been also shown that Si NPs can increase the PL emission from *MAPbI$_3$ perovskites*[86]. **Figure 6**(c) demonstrates the PL intensity spectra from hybrid Si NP & MAPbI$_3$ perovskite system (red curve) and from reference MAPbI$_3$ perovskite (black curve). 50% enhancement of PL intensity from the perovskite layer with Si RDNs and 200% enhancement for a nanoimprinted metasurface with Si NPs on top have been experimentally achieved owing to reduced non-radiative channels for energy dissipation in the presence of Si NPs and the Purcell effect.

A monolayer of *transition metal dichalcogenide* (TMDC)[263–265] is formed by a hexagonal layer of transition metal atoms (Mo, W) hosted between two hexagonal lattices of chalcogenide atoms (S, Se). Electronically, TMDCs behave as two-dimensional semiconductors with their bandgaps lying in the visible range. In the atomic monolayer limit these materials are particularly interesting, when their bandgap becomes direct enabling enhanced interaction of the dipole transition with light[266]. Due to their monolayer composition, high oscillator strength, and capabilities for tuning, these materials are a very promising class of materials in the context of various optoelectronic applications, such as photodetection and light harvesting[267], phototransistors and modulation[268], light-emitting diodes[269], and lasers[270]. In Ref.[271] the issue of resonance coupling in single Si NP-monolayer TMDC (WS$_2$) heterostructures has been addressed, **Figure 6**(d). A transition from weak to strong coupling regimes with coupling constant exceeding 200 meV for a Si NP covered by a monolayer WS$_2$ at the magnetic optical Mie resonance has been demonstrated. This effect has been achieved by changing the surrounding dielectric material from air to water. The ability of such system to tunable resonance coupling has been experimentally realized for spherical Si NPs arranged on a WS$_2$ flake. The coupling constant increases from 24.8 meV up to 43.4 meV by replacing air by water.

A variety of RDNs have been recently proposed and investigated for multiple applications. Their ability to localize light at the nanoscale and to boost light-matter interaction within different excitonic systems in both weak and strong coupling regimes have been demonstrated. The next two Sections are devoted to a review of the spectroscopy and biosensing applications of the proposed RDNs.

## C. Spectroscopy with resonant dielectric nanoantennas

Resonant NPs and nanostructures have been proven to be powerful tools for enhancing light-matter interactions due to their ability to enhance and localize optical energy at the nanoscale, boosting PL spectra through the Purcell effect and increasing the molecules excitation rates (see Section A). Two spectroscopy techniques based on these effects have been proposed: *surface enhanced Raman scattering* (SERS)[154,272–278] and *enhanced fluorescence spectroscopy*[21,169,279–281].

Despite plasmonics has demonstrated tremendous success in these phenomena[282,283], the concept of all-dielectric nanophotonics can also serve as a platform for high-effective detection scheme, even with single-molecule sensitivity. First of all, as we already mentioned in the previous section, low Ohmic losses in RDNs prevent parasitic heating of the analyzed objects[70,87,178], and second, high radiative part of the Purcell factor and directivity improve signal extraction[81]. **Figures 7**(a)-(d) show the enhancement of radiative decay rate and quantum efficiency (see Eq. (13)) of an electric dipolar emitter positioned between two Si (a), (b) and Au (c), (d) nanospheres for longitudinal and transverse dipole orientations. It can be seen that the Si nanoantennas have the quantum efficiency exceeding that of the Au nanoantennas[178]. Moreover, this good enhancement performance is achieved in combination with low heat generation. **Figures 7**(e)-(g) show the measured temperature increase on Si and Au nanoantennas when excited at the dipolar resonance. Notably, Si nanoantennas show a nearly constant temperature while the Au nanoantennas lead to a temperature increase of more than 100°C in the nanogap[70]. This allows to conduct spectroscopy measurements near nanoantennas at a constant temperature over a wide range of excitation powers.

**Figure 8**(a) shows experimental results of all-dielectric nanoantennas for surface enhanced Raman scattering (SERS) at the single-nanostructure level for the ensemble scenario (many molecules)[70]. The normalized SERS intensity map of nine similar disks-dimer silicon nanoantennas is shown in **Figure 8**(a), exhibiting the SERS signal increase on the nanoantennas. Such nanostructures exhibit high near-field enhancements within a 20-nm gap at the near IR wavelengths (see **Figure 8**(b) for the numerically calculated near-field map). Thanks to this near-field enhancement at a nanoscale volume, this all-dielectric nanoantenna

can enhance the Raman scattering of a polymer thin film by a factor of $10^3$, in excellent agreement with the theoretical expectation, as shown in **Figure 8**(c). Moreover, the authors performed molecular thermometry measurements to demonstrate that the dielectric nanoantennas produce ultra-low heating when illuminating at their resonance wavelength (results shown in **Figures 7**(e)-(g)), thus overcoming one of the main drawbacks of traditional plasmonic materials such as gold.

In addition to the SERS capabilities combined with low Ohmic losses, all-dielectric nanoantennas can be used efficiently for fluorescence enhancement. Importantly, due to the absence of quenching and of free carriers, which are inherent to metallic NPs, the photoluminescence experiments with dielectric NPs do not require any dielectric spacer. In Ref.[90] the enhancement in photoluminescence signal from Rhodamine B molecules up to 200 times with just single subwavelength dielectric (Si) NPs without any dielectric spacers has been demonstrated [see **Figure 6**(a)].

As in plasmonics, higher photoluminescence enhancement factors require more complicated structures like dimers, trimers and oligomers, which have been theoretically and experimentally studied recently[88,123,240,256,259,284–286]. In **Figure 9** we summarize the results obtained so far in many molecule fluorescence enhancement using all-dielectric dimer nanoantennas, from Refs.[70,89]. In both cases, the experimental scheme is similar: dielectric disk dimers with a nanometric gap are fabricated on a substrate and then covered with a polymer thin-film with embedded fluorescent molecules, allowing a controlled and uniform distribution of molecules on the nanoantennas, **Figure 9**(a). To excite and detect the enhanced fluorescence signal, an epi-confocal configuration was used. The main difference in the samples is the material used: while the first uses silicon (Si) dimers, the second uses gallium phosphide (GaP) dimers. Note that in both cases the molecules can be in direct contact with the nanostructure (*no spacers were used*) possible due to the reduced quenching effect of dielectric nanoantennas. The reported maximum enhancement factor for the Si dimers is $2 \cdot 10^3$ while for the GaP dimers is $3.6 \cdot 10^3$. The reported higher enhancement can be explained by the higher near-field intensity in the gap for the GaP nanoantennas and the better emission enhancement due to the nearly zero losses in the visible for this material[89]. To prove this fact further, the authors measured a fluorescence lifetime reduce of at least 22 times, limited by the experimental setup.

Recently, fluorescence enhancement using dielectric nanoantennas was pushed to the *single-molecule limit*[88]. A dimer of silicon discs with a 20-nm gap and a broad dipolar resonance at the visible was used to obtain a x270 fluorescence brightness increase for single molecules of crystal violet, over performing gold structures with similar gaps. The experimental approach for this experiment is illustrated in **Figure 10**(a): fluorescent molecules can diffuse around the nanostructure while the structure is excited resonantly and fluorescence time traces are recorded. **Figure 10**(b) shows the fluorescence time trace molecules diffusing around such a silicon dimer with 20 nm gap, exhibiting clear bursts that correspond to single-molecule enhancement events: a molecule diffuses into the hot spot at the nanogap and gets enhanced, emitting many photons in a short time, creating the burst seen in the time trace. By comparing the intensity recorded in the burst with the unenhanced molecular brightness the enhancement factor has been experimentally obtained[28,70,88].

The recorded fluorescence intensity time traces also contain information about the dynamics of the molecules in the solution, which can be exploited to extract information about the reduced detection volume of the nanogap antennas by *fluorescence correlation spectroscopy* (FCS) analysis[21]. **Figure 10**(c) shows the obtained correlation curve and (d) reports the obtained detection volumes in comparison with gold dimers with similar gaps. Note that the measured reduced volume can be as small as $\lambda^3/1800$, a value that is like the one obtained in a previous work in Au nanodimers. Next in the analysis, the authors measured the fluorescence lifetime to estimate the enhancement in the radiative rate of the molecule. In the enhanced case, they observed a new short component in the lifetime curve of 150 ps, shown in **Figure 10**(e). This value leads to a 15-fold increase in the radiative rate, considering the experimental conditions used. Finally, the authors compare the obtained overall enhancement values for two dyes with different quantum yield and observed an excellent agreement with the theoretical calculations, both for 20 and 30 nm gap, as evidenced in **Figure 10**(f).

Altogether, these results highlight the interesting opportunities offered by all-dielectric nanoantennas for fluorescence enhancement. Due to the remarkably low losses, we envision further applications in fluorescence enhancement of slow emitters such as rare earth ions with implications in lighting technology or for the creation of fast single-photon sources.

### D. Biosensing with resonant dielectric nanoparticles

The study of biological analytes through optical and nanophotonic structures (mainly metallic-based ones) has been an actively developing research area for a few decades[288–290]. The operation principle of the vast majority of proposed biosensors is based on the dependence of the plasmonic resonance of metal NPs on their environment, where the NPs are hosted[246,291–293]. Namely, the plasmonic resonance of a metal NP shifts if the refractive index of the surrounding dielectric changes (see **Figure 2**(e)). Therefore, if one activates the NPs by specific linker molecules that are targeted on a certain substance that should be detected, it will cause the measurable shift of the resonance when the substance appears in the NP environment. As we already mentioned in Section A., the figure of merit (FoM) of plasmonic sensors is commonly defined as the resonance shift (S) upon a change in the refractive index of the surrounding dielectric normalized by the resonance line width ($\Delta\lambda$), see Eq. (7). Thus, narrow resonances improve the FoM of plasmonic sensors[288]. Specific applications of SPR-based biosensing that can be found in the literature include detection of contaminants in food and water[294], immunoassays[295], DNA-protein interactions[296], and detection and monitoring of antigen-antibody binding events in real time[297].

However, the sensing performance of plasmonic nanostructures is limited by their conductive and absorptive nature due to free electrons. The highly absorptive nature of metals inducing undesirable local heating[147,298], which can lead to denaturation of molecules and hinder the development for *in vivo* sensing[299]. Very recently, it has been proven that RDNs may become a powerful platform for biological and medical applications including optical biosensors[260,300,301]. On the one hand, high-index RDNs possess low dissipative losses, and their heating at ordinary intensities of laser excitation is sufficiently lower[70,178]. On the other hand, RDNs exhibit both electric and magnetic dipole Mie resonances, which offers more ability to detect the changes in ambient conditions -- altering of scattering pattern. It allows one to use them in biosensing applications by similar way as it is done for plasmon sensors.

In Ref.[300] both sensing principles, namely (i) the intensity variations in the transmitted or backscattered radiation by the RDNs at the magnetic dipole resonance and (ii) the changes in the radiation pattern at the frequency that satisfies Kerker condition (see Section B.) of near-zero forward radiation have been theoretically proposed. This idea has been experimentally demonstrated by Bontempi et al. in Ref.[157] in the system of 2D array of Si nanoparticles (metasurface) sensitive to streptavidin. To achieve this functionality, the Si metasurface, previously fabricated by e-beam lithography, has been activated generating hydroxyl radicals (--OH) on the surface of nanodisks. This surface activation process was possible because silicon has a thin layer of native oxide after fabrication[70,302]. Then, to perform linker functionalization, the silane-polyethylene glycol (PEG)-biotin molecules as the linking medium between the Si nanodisks and streptavidin have been used, **Figure 11**(a). When the streptavidin bounds to the biotin on the linker molecule using the binding sites it changes the metasurface transmittance spectra, **Figure 11**(b). The collective resonance (dip around 1.5 μm) is very sensitive to the streptavidin in the biological analyte and even $10^{-5}$ M concentrations cause its red-shifting by 10 nm, which can be resolved by standard optical means. After the removal of the streptavidin (purple curve), the resonance position returns to the position prior to the application of streptavidin (the biotin-coated case). **Figure 11**(c) demonstrates the color map of the transmittance spectra of nanodisk arrays measured with vertically polarized light for different streptavidin concentrations. The resonance at around 1488 nm red shifts as the concentration of streptavidin increases and can be seen by the ascending blue area denoting the transmittance dip. **Figure 11**(d) shows the sensitivity of this approach. The results demonstrate the streptavidin detection limit of $10^{-10}$, which is two orders of magnitudes better than existing nanosensors[303]. Similar approach has been proposed in Ref.[304] for the periodic silicon nanowire-based arrays. It has been demonstrated that by functionalizing the surface of Si RDNs with a graphene monolayer, such structures can be used to optically detect low-concentration surface adsorption events. The specific label-free detection limit using immunoglobulin G protein (IgG) is found to be approximately 300 pM with an extracted maximum sensor resonance shift of 5.42 nm.

The RDNs provide wide opportunities to design more complex systems possessing Fano resonances[65,123,125,259,284] (see Section A). It has been demonstrated that at the Fano resonance frequency the near-field of RDNs increases to levels even larger than of a single constituting element, providing an additional powerful tool to enhancing light-matter interaction effects. Therefore, the Fano resonance can be utilized to improve the sensor FoM, which has been realized in Ref.[260], where a FoM of 103 for Si metasurface has been reported. We note that this value is significantly higher than the best FoMs obtained with Fano resonant plasmonic sensors, that are on the order of 23.[305]

The application of high-index RDNs for biosensing has been further developed in Ref.[306], where it has been combined with a microfluidics system, thus realizing the lab-on-a-chip platform enables detecting low concentrations of prostate specific antigen (PSA) cancer marker in human serum. **Figure 11**(e) shows the picture of a microfluidic system with eight channels, which include optically resonant silicon nanodisk arrays of different dimensions and lattice parameter. The operation principle of the system is also based on the dependence of RDN collective resonances on the environment conditions. Namely, the increased effective refractive index at the sensor vicinity resulting from the binding steps causes the resonances of the nanodisk arrays to redshift enabling an accurate monitoring of the adsorption and binding kinetics of the molecules, **Figure 11**(f). Moreover, it has been demonstrated by direct comparison that the operation performance of RDN-based devices far exceeds its Au counterpart, including higher sensitivity and higher stability in solution.

## Summary and Outlook

In this paper, we have reviewed the state-of-the-art applications of optically resonant high-index dielectric nanostructures as a multifunctional platform for light-matter interactions, especially focusing on spectroscopy and biosensing. We have given a comprehensive survey of surface enhanced Raman scattering, fluorescence spectroscopy, biosensing and lab-on-chip technologies based on RDNs. We have accompanied the work by a theoretical background underlying the operation principles of these techniques, with an overview of specific resonant dielectric nanostructure and hybrid excitonic systems.

To summarize all results presented in the Review, we may conclude that resonant dielectric nanostructures have many advantages compared to plasmonic ones, including low energy dissipation into heat, resonant enhancement of magnetic fields in dielectric nanoparticles and bring novel functionalities in both spectroscopy and sensing applications. **Table 1** summarizes the presented results with sensitivity and FoM for different nanostrucutres used for biosensing. More details about plasmonic-based biosensing can be found in the literature[307], here we reproduced only a small subset for comparison with RDNs. We may conclude saying that the most sensitive sensor, from the point of view of FoM, presented to date has been proposed in Ref.[260]. This sensor consists of a Si metasurface with Fano-resonant double-gap split-ring elements. The Fano resonance provides a narrow peak in the transmittance spectra, which is sensitive to the surrounding dielectric variations. The sensitivity of the proposed RDNs does not exceed the sensitivity of the best plasmonic devices realized up to date. However, the trend allows one to expect the emergence of more sensitive RDN-based devices in the near future.

Thus, despite impressive and encouraging results in this area, consensus about the best performing designs has not been formed yet. This is a young research area where more studies and experiments are needed. We envision several future research directions in this context.

First, it has been recently demonstrated that second-harmonic generation from dielectric nanoparticles made of bio-compatible hybrid perovskites (barium titanate, $BaTiO_3$) can be significantly enhanced at the magnetic Mie resonance[308]. This opens a way to use such particles for different kind of bioimaging applications, where they can be efficiently used as second-harmonic imaging probes in in vitro and in vivo tissues[309,310]. Moreover, an additional enhancement of second-harmonic generation from $BaTiO_3$ nanoparticles has been achieved by additional coupling to a surface plasmon resonance of a gold nanoparticle at the pumping frequency[311].

Second, the concept of all-dielectric nanophotonics can be a new approach to provide effective detection of molecular events beyond the diffraction limit, while revealing analyte conformation and its local temperature in real time. It has been shown that silicon nanoparticles possess resonantly enhanced inherent Raman response[121]. This effect can be potentially utilized for simultaneous detection of a nanoscale molecular event (protein unfolding) and control of the local temperature in real time. The high sensitivity and sub-diffraction spatial resolution can be achieved due to resonant properties of silicon nanoantennas, whereas precise nanothermometry is caused by the effect of heating on the Raman scattering signal.

Finally, as discussed above, increasing of the resonance quality factor may make the sensor more sensitive. In this regard, it is useful to note that almost non-radiating resonant states called bound states in the continuum (BICs) have been recently discovered in photonics[312–315]. These are localized states in the continuous spectrum of an electromagnetically open resonant (nano)structure. An existence of these states allows designing photonic resonators with diverging and potentially *infinite lifetimes and quality factors*, leading to many vital applications, including biosensors. It has been recently demonstrated that RDNs are a powerful platform for designing BIC states[316,317] which can be explored for advanced spectroscopy and biosensor applications.

## Authors information

The authors declare no conflict of interest.

## Acknowledgements

This work was also partially supported by the Air Force Office of Scientific Research and by the Welch Foundation with grant No, F-1802.

Received: ((will be filled in by the editorial staff))
Revised: ((will be filled in by the editorial staff))
Published online: ((will be filled in by the editorial staff))

## References

(1) Bharadwaj, P.; Deutsch, B.; Novotny, L. Optical Antennas. *Adv. Opt. Photonics* **2009**, *1* (3), 438.

(2) Novotny, L.; van Hulst, N. Antennas for Light. *Nat. Photonics* **2011**, *5* (2), 83–90.

(3) Krasnok, A. E.; Maksymov, I. S.; Denisyuk, A. I.; Belov, P. A.; Miroshnichenko, A. E.; Simovski, C. R.; Kivshar, Y. S. Optical Nanoantennas. *Physics-Uspekhi* **2013**, *56* (6), 539–564.

(4) Liu, N.; Wen, F.; Zhao, Y.; Wang, Y.; Nordlander, P.; Halas, N. J.; Alù, A. Individual Nanoantennas Loaded with Three-Dimensional Optical Nanocircuits. *Nano Lett.* **2013**, *13* (1), 142–147.

(5) Alu, A.; Engheta, N. Theory, Modeling and Features of Optical Nanoantennas. *IEEE*

*Trans. Antennas Propag.* **2013**, *61* (4), 1508–1517.

(6) Monticone, F.; Alù, A. Metamaterials and Plasmonics: From Nanoparticles to Nanoantenna Arrays, Metasurfaces, and Metamaterials. *Chinese Phys. B* **2014**, *23* (4), 47809.

(7) Koenderink, A. F.; Alu, A.; Polman, A. Nanophotonics: Shrinking Light-Based Technology. *Science (80-. ).* **2015**, *348* (6234), 516–521.

(8) Monticone, F.; Alù, A. Metamaterial, Plasmonic and Nanophotonic Devices. *Reports Prog. Phys.* **2017**, *80* (3), 36401.

(9) Koenderink, A. F. Single-Photon Nanoantennas. *ACS Photonics* **2017**, *4* (4), 710–722.

(10) Dionne, J. A.; Baldi, A.; Baum, B.; Ho, C.-S.; Janković, V.; Naik, G. V.; Narayan, T.; Scholl, J. A.; Zhao, Y. Localized Fields, Global Impact: Industrial Applications of Resonant Plasmonic Materials. *MRS Bull.* **2015**, *40* (12), 1138–1145.

(11) Anger, P.; Bharadwaj, P.; Novotny, L. Enhancement and Quenching of Single-Molecule Fluorescence. *Phys. Rev. Lett.* **2006**, *96* (11), 113002.

(12) Kühn, S.; Håkanson, U.; Rogobete, L.; Sandoghdar, V. Enhancement of Single-Molecule Fluorescence Using a Gold Nanoparticle as an Optical Nanoantenna. *Phys. Rev. Lett.* **2006**, *97* (1), 17402.

(13) Biagioni, P.; Huang, J.-S.; Hecht, B. Nanoantennas for Visible and Infrared Radiation. *Reports Prog. Phys.* **2011**, *75* (2), 24402.

(14) Marquier, F.; Sauvan, C.; Greffet, J.-J. Revisiting Quantum Optics with Surface Plasmons and Plasmonic Resonators. *ACS Photonics* **2017**, *4* (9), acsphotonics.7b00475.

(15) Kinkhabwala, A.; Yu, Z.; Fan, S.; Avlasevich, Y.; Müllen, K.; Moerner, W. E. Large Single-Molecule Fluorescence Enhancements Produced by a Bowtie Nanoantenna. *Nat. Photonics* **2009**, *3* (11), 654–657.

(16) Bharadwaj, P.; Beams, R.; Novotny, L. Nanoscale Spectroscopy with Optical Antennas.

*Chem. Sci.* **2011**, *2* (1), 136–140.

(17) Acuna, G. P.; Moller, F. M.; Holzmeister, P.; Beater, S.; Lalkens, B.; Tinnefeld, P. Fluorescence Enhancement at Docking Sites of DNA-Directed Self-Assembled Nanoantennas. *Science (80-. ).* **2012**, *338* (6106), 506–510.

(18) Busson, M. P.; Rolly, B.; Stout, B.; Bonod, N.; Bidault, S. Accelerated Single Photon Emission from Dye Molecule-Driven Nanoantennas Assembled on DNA. *Nat. Commun.* **2012**, *3*, 962.

(19) Punj, D.; Mivelle, M.; Moparthi, S. B.; van Zanten, T. S.; Rigneault, H.; van Hulst, N. F.; García-Parajó, M. F.; Wenger, J. A Plasmonic "Antenna-in-Box" Platform for Enhanced Single-Molecule Analysis at Micromolar Concentrations. *Nat. Nanotechnol.* **2013**, *8* (7), 512–516.

(20) Khatua, S.; Paulo, P. M. R.; Yuan, H.; Gupta, A.; Zijlstra, P.; Orrit, M. Resonant Plasmonic Enhancement of Single-Molecule Fluorescence by Individual Gold Nanorods. *ACS Nano* **2014**, *8* (5), 4440–4449.

(21) Punj, D.; Regmi, R.; Devilez, A.; Plauchu, R.; Moparthi, S. B.; Stout, B.; Bonod, N.; Rigneault, H.; Wenger, J. Self-Assembled Nanoparticle Dimer Antennas for Plasmonic-Enhanced Single-Molecule Fluorescence Detection at Micromolar Concentrations. *ACS Photonics* **2015**, *2* (8), 1099–1107.

(22) Akselrod, G. M.; Weidman, M. C.; Li, Y.; Argyropoulos, C.; Tisdale, W. A.; Mikkelsen, M. H. Efficient Nanosecond Photoluminescence from Infrared PbS Quantum Dots Coupled to Plasmonic Nanoantennas. *ACS Photonics* **2016**, *3* (10), 1741–1746.

(23) Trofymchuk, K.; Reisch, A.; Didier, P.; Fras, F.; Gilliot, P.; Mely, Y.; Klymchenko, A. S. Giant Light-Harvesting Nanoantenna for Single-Molecule Detection in Ambient Light. *Nat. Photonics* **2017**, *11* (10), 657–663.

(24) Kress, S. J. P.; Cui, J.; Rohner, P.; Kim, D. K.; Antolinez, F. V.; Zaininger, K.-A.;

Jayanti, S. V.; Richner, P.; McPeak, K. M.; Poulikakos, D.; Norris, D. J. Colloidal-Quantum-Dot Spasers and Plasmonic Amplifiers. *Sci. Adv.* **2016**, *3* (9), e1700688.

(25) Akselrod, G. M.; Argyropoulos, C.; Hoang, T. B.; Ciracì, C.; Fang, C.; Huang, J.; Smith, D. R.; Mikkelsen, M. H. Probing the Mechanisms of Large Purcell Enhancement in Plasmonic Nanoantennas. *Nat. Photonics* **2014**, *8* (11), 835–840.

(26) Straubel, J.; Filter, R.; Rockstuhl, C.; Słowik, K. Plasmonic Nanoantenna Based Triggered Single-Photon Source. *Phys. Rev. B* **2016**, *93* (19), 195412.

(27) Hoang, T. B.; Akselrod, G. M.; Mikkelsen, M. H. Ultrafast Room-Temperature Single Photon Emission from Quantum Dots Coupled to Plasmonic Nanocavities. *Nano Lett.* **2016**, *16* (1), 270–275.

(28) Chikkaraddy, R.; de Nijs, B.; Benz, F.; Barrow, S. J.; Scherman, O. A.; Rosta, E.; Demetriadou, A.; Fox, P.; Hess, O.; Baumberg, J. J. Single-Molecule Strong Coupling at Room Temperature in Plasmonic Nanocavities. *Nature* **2016**, *535* (7610), 127–130.

(29) Baranov, D. G.; Wersäll, M.; Cuadra, J.; Antosiewicz, T. J.; Shegai, T. Novel Nanostructures and Materials for Strong Light-Matter Interactions. *ACS Photonics* **2017**, *1*, acsphotonics.7b00674.

(30) Dykman, L.; Khlebtsov, N. Gold Nanoparticles in Biomedical Applications: Recent Advances and Perspectives. *Chem. Soc. Rev.* **2012**, *41* (6), 2256–2282.

(31) Huang, J. A.; Zhang, Y. L.; Ding, H.; Sun, H. B. SERS-Enabled Lab-on-a-Chip Systems. *Adv. Opt. Mater.* **2015**, *3* (5), 618–633.

(32) Gwo, S.; Wang, C. Y.; Chen, H. Y.; Lin, M. H.; Sun, L.; Li, X.; Chen, W. L.; Chang, Y. M.; Ahn, H. Plasmonic Metasurfaces for Nonlinear Optics and Quantitative SERS. *ACS Photonics* **2016**, *3* (8), 1371–1384.

(33) Alù, A.; Engheta, N. Wireless at the Nanoscale: Optical Interconnects Using Matched Nanoantennas. *Phys. Rev. Lett.* **2010**, *104* (21), 213902.

(34) Alù, A.; Engheta, N. Input Impedance, Nanocircuit Loading, and Radiation Tuning of

Optical Nanoantennas. *Phys. Rev. Lett.* **2008**, *101* (4), 43901.

(35) Curto, A. G.; Volpe, G.; Taminiau, T. H.; Kreuzer, M. P.; Quidant, R.; van Hulst, N. F. Unidirectional Emission of a Quantum Dot Coupled to a Nanoantenna. *Science (80-. ).* **2010**, *329* (5994), 930–933.

(36) Aouani, H.; Mahboub, O.; Bonod, N.; Devaux, E.; Popov, E.; Rigneault, H.; Ebbesen, T. W.; Wenger, J. Bright Unidirectional Fluorescence Emission of Molecules in a Nanoaperture with Plasmonic Corrugations. *Nano Lett.* **2011**, *11* (2), 637–644.

(37) Moiseev, E. I.; Kryzhanovskaya, N.; Polubavkina, Y. S.; Maximov, M. V.; Kulagina, M. M.; Zadiranov, Y. M.; Lipovskii, A. A.; Mukhin, I. S.; Mozharov, A. M.; Komissarenko, F. E.; Sadrieva, Z. F.; Krasnok, A. E.; Bogdanov, A. A.; Lavrinenko, A. V.; Zhukov, A. E. Light Outcoupling from Quantum Dot-Based Microdisk Laser via Plasmonic Nanoantenna. *ACS Photonics* **2017**, *4* (2), 275–281.

(38) Alù, A.; Engheta, N. Tuning the Scattering Response of Optical Nanoantennas with Nanocircuit Loads. *Nat. Photonics* **2008**, *2* (5), 307–310.

(39) Eghlidi, H.; Lee, K. G.; Chen, X. W.; Götzinger, S.; Sandoghdar, V. Resolution and Enhancement in Nanoantenna-Based Fluorescence Microscopy. *Nano Lett.* **2009**, *9* (12), 4007–4011.

(40) Van Zanten, T. S.; Lopez-Bosque, M. J.; Garcia-Parajo, M. F. Imaging Individual Proteins and Nanodomains on Intact Cell Membranes with a Probe-Based Optical Antenna. *Small* **2010**, *6* (2), 270–275.

(41) Mivelle, M.; Van Zanten, T. S.; Garcia-Parajo, M. F. Hybrid Photonic Antennas for Subnanometer Multicolor Localization and Nanoimaging of Single Molecules. *Nano Lett.* **2014**, *14* (8), 4895–4900.

(42) Zhang, S.; Bao, K.; Halas, N. J.; Xu, H.; Nordlander, P. Substrate-Induced Fano Resonances of a Plasmonic Nanocube: A Route to Increased-Sensitivity Localized Surface Plasmon Resonance Sensors Revealed. *Nano Lett.* **2011**, *11* (4), 1657–1663.

(43) Pellegrotti, J. V.; Cortés, E.; Bordenave, M. D.; Caldarola, M.; Kreuzer, M. P.; Sanchez, A. D.; Ojea, I.; Bragas, A. V.; Stefani, F. D. Plasmonic Photothermal Fluorescence Modulation for Homogeneous Biosensing. *ACS Sensors* **2016**, *1* (11), 1351–1357.

(44) Atwater, H. A.; Polman, A. Plasmonics for Improved Photovoltaic Devices. *Nat. Mater.* **2010**, *9* (10), 865–865.

(45) Spinelli, P.; Ferry, V. E.; van de Groep, J.; van Lare, M.; Verschuuren, M. a; Schropp, R. E. I.; Atwater, H. a; Polman, a. Plasmonic Light Trapping in Thin-Film Si Solar Cells. *J. Opt.* **2012**, *14* (2), 24002.

(46) Knight, M. W.; Sobhani, H.; Nordlander, P.; Halas, N. J. Photodetection with Active Optical Antennas. *Science (80-. ).* **2011**, *332* (6030), 702–704.

(47) Hirsch, L. R.; Stafford, R. J.; Bankson, J. A.; Sershen, S. R.; Rivera, B.; Price, R. E.; Hazle, J. D.; Halas, N. J.; West, J. L. Nanoshell-Mediated near-Infrared Thermal Therapy of Tumors under Magnetic Resonance Guidance. *Proc. Natl. Acad. Sci. U. S. A.* **2003**, *100* (23), 13549–13554.

(48) Kim, K.; Kim, J. H.; Park, H.; Kim, Y. S.; Park, K.; Nam, H.; Lee, S.; Park, J. H.; Park, R. W.; Kim, I. S.; Choi, K.; Kim, S. Y.; Park, K.; Kwon, I. C. Tumor-Homing Multifunctional Nanoparticles for Cancer Theragnosis: Simultaneous Diagnosis, Drug Delivery, and Therapeutic Monitoring. *J. Control. Release* **2010**, *146* (2), 219–227.

(49) Naik, G. V.; Shalaev, V. M.; Boltasseva, A. Alternative Plasmonic Materials: Beyond Gold and Silver. *Adv. Mater.* **2013**, *25* (24), 3264–3294.

(50) West, P. R.; Ishii, S.; Naik, G. V.; Emani, N. K.; Shalaev, V. M.; Boltasseva, A. Searching for Better Plasmonic Materials. *Laser Photonics Rev.* **2010**, *4* (6), 795–808.

(51) Boltasseva, A.; Atwater, H. A. Low-Loss Plasmonic Metamaterials. *Science (80-. ).* **2011**, *331* (6015), 290–291.

(52) Tassin, P.; Koschny, T.; Kafesaki, M.; Soukoulis, C. M. A Comparison of Graphene, Superconductors and Metals as Conductors for Metamaterials and Plasmonics. *Nat.*

*Photonics* **2012**, *6* (4), 259–264.

(53) Khurgin, J. B. How to Deal with the Loss in Plasmonics and Metamaterials. *Nat. Nanotechnol.* **2015**, *10* (1), 2–6.

(54) Khurgin, J. B. Ultimate Limit of Field Confinement by Surface Plasmon Polaritons. *Faraday Discuss.* **2015**, *178*, 109–122.

(55) Khurgin, J. B. Replacing Noble Metals with Alternative Materials in Plasmonics and Metamaterials: How Good an Idea? *Philos. Trans. R. Soc. A Math. Phys. Eng. Sci.* **2017**, *375* (2090), 20160068.

(56) Zhao, Q.; Zhou, J.; Zhang, F.; Lippens, D. Mie Resonance-Based Dielectric Metamaterials. *Mater. Today* **2009**, *12* (12), 60–69.

(57) Evlyukhin, A. B.; Reinhardt, C.; Seidel, A.; Luk'Yanchuk, B. S.; Chichkov, B. N. Optical Response Features of Si-Nanoparticle Arrays. *Phys. Rev. B - Condens. Matter Mater. Phys.* **2010**, *82* (4), 45404.

(58) Evlyukhin, A. B.; Novikov, S. M.; Zywietz, U.; Eriksen, R. L.; Reinhardt, C.; Bozhevolnyi, S. I.; Chichkov, B. N. Demonstration of Magnetic Dipole Resonances of Dielectric Nanospheres in the Visible Region. *Nano Lett.* **2012**, *12* (7), 3749–3755.

(59) Kuznetsov, A. I.; Miroshnichenko, A. E.; Fu, Y. H.; Zhang, J.; Luk'yanchuk, B. Magnetic Light. *Sci. Rep.* **2012**, *2* (1), 492.

(60) Zywietz, U.; Evlyukhin, A. B.; Reinhardt, C.; Chichkov, B. N. Laser Printing of Silicon Nanoparticles with Resonant Optical Electric and Magnetic Responses. *Nat. Commun.* **2014**, *5*, 3402.

(61) Evlyukhin, A. B.; Eriksen, R. L.; Cheng, W.; Beermann, J.; Reinhardt, C.; Petrov, A.; Prorok, S.; Eich, M.; Chichkov, B. N.; Bozhevolnyi, S. I. Optical Spectroscopy of Single Si Nanocylinders with Magnetic and Electric Resonances. *Sci. Rep.* **2014**, *4*, 4126.

(62) Krasnok, A.; Makarov, S.; Petrov, M.; Savelev, R.; Belov, P.; Kivshar, Y. Towards

All-Dielectric Metamaterials and Nanophotonics. In *Proc. SPIE*; Kuzmiak, V., Markos, P., Szoplik, T., Eds.; 2015; Vol. 9502, p 950203.

(63) Savelev, R. S.; Makarov, S. V.; Krasnok, A. E.; Belov, P. A. From Optical Magnetic Resonance to Dielectric Nanophotonics (A Review). *Opt. Spectrosc.* **2015**, *119* (4), 551–568.

(64) Kuznetsov, A. I.; Miroshnichenko, A. E.; Brongersma, M. L.; Kivshar, Y. S.; Luk'yanchuk, B. Optically Resonant Dielectric Nanostructures. *Science (80-. ).* **2016**, *354* (6314), aag2472.

(65) Decker, M.; Staude, I. Resonant Dielectric Nanostructures: A Low-Loss Platform for Functional Nanophotonics. *J. Opt.* **2016**, *18* (10), 103001.

(66) Staude, I.; Schilling, J. Metamaterial-Inspired Silicon Nanophotonics. *Nat. Photonics* **2017**, *11* (5), 274–284.

(67) Krasnok, A.; Tymchenko, M.; Alù, A. Nonlinear Metasurfaces: A Paradigm Shift in Nonlinear Optics. *Mater. Today* **2017**, *0* (0), 1–60.

(68) Baranov, D. G.; Zuev, D. A.; Lepeshov, S. I.; Kotov, O. V.; Krasnok, A. E.; Evlyukhin, A. B.; Chichkov, B. N. All-Dielectric Nanophotonics: The Quest for Better Materials and Fabrication Techniques. *Optica* **2017**, *4* (7), 814.

(69) Devilez, A.; Zambrana-Puyalto, X.; Stout, B.; Bonod, N. Mimicking Localized Surface Plasmons with Dielectric Particles. *Phys. Rev. B* **2015**, *92* (24), 241412.

(70) Caldarola, M.; Albella, P.; Cortés, E.; Rahmani, M.; Roschuk, T.; Grinblat, G.; Oulton, R. F.; Bragas, A. V.; Maier, S. A. Non-Plasmonic Nanoantennas for Surface Enhanced Spectroscopies with Ultra-Low Heat Conversion. *Nat. Commun.* **2015**, *6*, 7915.

(71) Shcherbakov, M. R.; Shorokhov, A. S.; Neshev, D. N.; Hopkins, B.; Staude, I.; Melik-Gaykazyan, E. V.; Ezhov, A. A.; Miroshnichenko, A. E.; Brener, I.; Fedyanin, A. A.; Kivshar, Y. S. Nonlinear Interference and Tailorable Third-Harmonic Generation from Dielectric Oligomers. *ACS Photonics* **2015**, *2* (5), 578–582.

(72) Makarov, S. V.; Tsypkin, A. N.; Voytova, T. A.; Milichko, V. A.; Mukhin, I. S.; Yulin, A. V.; Putilin, S. E.; Baranov, M. A.; Krasnok, A. E.; Morozov, I. A.; Belov, P. A. Self-Adjusted All-Dielectric Metasurfaces for Deep Ultraviolet Femtosecond Pulse Generation. *Nanoscale* **2016**, *8* (41), 17809–17814.

(73) Smirnova, D.; Kivshar, Y. S. Multipolar Nonlinear Nanophotonics. *Optica* **2016**, *3* (11), 1241.

(74) Camacho-Morales, R.; Rahmani, M.; Kruk, S.; Wang, L.; Xu, L.; Smirnova, D. A.; Solntsev, A. S.; Miroshnichenko, A.; Tan, H. H.; Karouta, F.; Naureen, S.; Vora, K.; Carletti, L.; De Angelis, C.; Jagadish, C.; Kivshar, Y. S.; Neshev, D. N. Nonlinear Generation of Vector Beams from AlGaAs Nanoantennas. *Nano Lett.* **2016**, *16* (11), 7191–7197.

(75) Carletti, L.; Locatelli, A.; Neshev, D.; De Angelis, C. Shaping the Radiation Pattern of Second-Harmonic Generation from AlGaAs Dielectric Nanoantennas. *ACS Photonics* **2016**, *3* (8), 1500–1507.

(76) Shibanuma, T.; Grinblat, G.; Albella, P.; Maier, S. A. Efficient Third Harmonic Generation from Metal–Dielectric Hybrid Nanoantennas. *Nano Lett.* **2017**, *17* (4), 2647–2651.

(77) Grinblat, G.; Li, Y.; Nielsen, M. P.; Oulton, R. F.; Maier, S. A. Enhanced Third Harmonic Generation in Single Germanium Nanodisks Excited at the Anapole Mode. *Nano Lett.* **2016**, *16* (7), 4635–4640.

(78) Kruk, S. S.; Camacho-Morales, R.; Xu, L.; Rahmani, M.; Smirnova, D. A.; Wang, L.; Tan, H. H.; Jagadish, C.; Neshev, D. N.; Kivshar, Y. S. Nonlinear Optical Magnetism Revealed by Second-Harmonic Generation in Nanoantennas. *Nano Lett.* **2017**, *17* (6), 3914–3918.

(79) Slobozhanyuk, A.; Mousavi, S. H.; Ni, X.; Smirnova, D.; Kivshar, Y. S.; Khanikaev, A. B. Three-Dimensional All-Dielectric Photonic Topological Insulator. *Nat. Photonics*

**2016**, *11* (2), 130–136.

(80) Kruk, S.; Slobozhanyuk, A.; Denkova, D.; Poddubny, A.; Kravchenko, I.; Miroshnichenko, A.; Neshev, D.; Kivshar, Y. Edge States and Topological Phase Transitions in Chains of Dielectric Nanoparticles. *Small* **2017**, *13* (11), 1603190.

(81) Krasnok, A. E.; Maloshtan, A.; Chigrin, D. N.; Kivshar, Y. S.; Belov, P. A. Enhanced Emission Extraction and Selective Excitation of NV Centers with All-Dielectric Nanoantennas. *Laser Photonics Rev.* **2015**, *9* (4), 385–391.

(82) Staude, I.; Khardikov, V. V.; Fofang, N. T.; Liu, S.; Decker, M.; Neshev, D. N.; Luk, T. S.; Brener, I.; Kivshar, Y. S. Shaping Photoluminescence Spectra with Magnetoelectric Resonances in All-Dielectric Nanoparticles. *ACS Photonics* **2015**, *2* (2), 172–177.

(83) Peter, M.; Hildebrandt, A.; Schlickriede, C.; Gharib, K.; Zentgraf, T.; Förstner, J.; Linden, S. Directional Emission from Dielectric Leaky-Wave Nanoantennas. *Nano Lett.* **2017**, *17* (7), 4178–4183.

(84) Rutckaia, V.; Heyroth, F.; Novikov, A.; Shaleev, M.; Petrov, M. I.; Schilling, J. Quantum Dot Emission Driven by Mie Resonances in Silicon Nanostructures. *Nano Lett.* **2017**, acs.nanolett.7b03248.

(85) Yuan, S.; Qiu, Xi.; Cui, C.; Zhu, L.; Wang, Y.; Li, Y.; Song, J.; Huang, Q.; Xia, J. Strong Photoluminescence Enhancement in All-Dielectric Fano Metasurface with High Quality Factor. *ACS Nano* **2017**, acsnano.7b04810.

(86) Tiguntseva, E.; Chebykin, A.; Ishteev, A.; Haroldson, R.; Balachandran, B.; Ushakova, E.; Komissarenko, F.; Wang, H.; Milichko, V.; Tsypkin, A.; Zuev, D.; Hu, W.; Makarov, S.; Zakhidov, A. Resonant Silicon Nanoparticles for Enhancement of Light Absorption and Photoluminescence from Hybrid Perovskite Films and Metasurfaces. *Nanoscale* **2017**, *5*, 3849–3853.

(87) Albella, P.; De La Osa, R. A.; Moreno, F.; Maier, S. A. Electric and Magnetic Field Enhancement with Ultralow Heat Radiation Dielectric Nanoantennas: Considerations for Surface-Enhanced Spectroscopies. *ACS Photonics* **2014**, *1* (6), 524–529.

(88) Regmi, R.; Berthelot, J.; Winkler, P. M.; Mivelle, M.; Proust, J.; Bedu, F.; Ozerov, I.; Begou, T.; Lumeau, J.; Rigneault, H.; García-Parajó, M. F.; Bidault, S.; Wenger, J.; Bonod, N. All-Dielectric Silicon Nanogap Antennas to Enhance the Fluorescence of Single Molecules. *Nano Lett.* **2016**, *16* (8), 5143–5151.

(89) Cambiasso, J.; Grinblat, G.; Li, Y.; Rakovich, A.; Cortés, E.; Maier, S. A. Bridging the Gap between Dielectric Nanophotonics and the Visible Regime with Effectively Lossless Gallium Phosphide Antennas. *Nano Lett.* **2017**, *17* (2), 1219–1225.

(90) Sugimoto, H.; Fujii, M. Colloidal Dispersion of Subquarter Micrometer Silicon Spheres for Low-Loss Antenna in Visible Regime. *Adv. Opt. Mater.* **2017**, *5* (17), 1700332.

(91) Ma, X.; James, A. R.; Hartmann, N. F.; Baldwin, J. K.; Dominguez, J.; Sinclair, M. B.; Luk, T. S.; Wolf, O.; Liu, S.; Doorn, S. K.; Htoon, H.; Brener, I. Solitary Oxygen Dopant Emission from Carbon Nanotubes Modified by Dielectric Metasurfaces. *ACS Nano* **2017**, *11* (6), 6431–6439.

(92) Krasnok, A. E.; Filonov, D. S.; Simovski, C. R.; Kivshar, Y. S.; Belov, P. A. Experimental Demonstration of Superdirective Dielectric Antenna. *Appl. Phys. Lett.* **2014**, *104* (13), 133502.

(93) Zuev, D. A.; Makarov, S. V.; Mukhin, I. S.; Milichko, V. A.; Starikov, S. V.; Morozov, I. A.; Shishkin, I. I.; Krasnok, A. E.; Belov, P. A. Fabrication of Hybrid Nanostructures via Nanoscale Laser-Induced Reshaping for Advanced Light Manipulation. *Adv. Mater.* **2016**, *28* (16), 3087–3093.

(94) Markovich, D.; Baryshnikova, K.; Shalin, A.; Samusev, A.; Krasnok, A.; Belov, P.; Ginzburg, P. Enhancement of Artificial Magnetism via Resonant Bianisotropy. *Sci. Rep.* **2016**, *6* (1), 22546.

(95) Li, J.; Verellen, N.; Vercruysse, D.; Bearda, T.; Lagae, L.; Van Dorpe, P. All-Dielectric

Antenna Wavelength Router with Bidirectional Scattering of Visible Light. *Nano Lett.* **2016**, *16* (7), 4396–4403.

(96) Shibanuma, T.; Matsui, T.; Roschuk, T.; Wojcik, J.; Mascher, P.; Albella, P.; Maier, S. A. Experimental Demonstration of Tunable Directional Scattering of Visible Light from All-Dielectric Asymmetric Dimers. *ACS Photonics* **2017**, *4* (3), 489–494.

(97) Garcia-Etxarri, A. Optical Polarization Möbius Strips on All-Dielectric Optical Scatterers. *ACS Photonics* **2017**, *4* (5), 1159–1164.

(98) Khaidarov, E.; Hao, H.; Paniagua-Domínguez, R.; Yu, Y. F.; Fu, Y. H.; Valuckas, V.; Yap, S. L. K.; Toh, Y. T.; Ng, J. S. K.; Kuznetsov, A. I. Asymmetric Nanoantennas for Ultrahigh Angle Broadband Visible Light Bending. *Nano Lett.* **2017**, acs.nanolett.7b02952.

(99) Noskov, R. E.; Krasnok, A. E.; Kivshar, Y. S. Nonlinear Metal-Dielectric Nanoantennas for Light Switching and Routing. *New J. Phys.* **2012**, *14* (9), 93005.

(100) Makarov, S.; Kudryashov, S.; Mukhin, I.; Mozharov, A.; Milichko, V.; Krasnok, A.; Belov, P. Tuning of Magnetic Optical Response in a Dielectric Nanoparticle by Ultrafast Photoexcitation of Dense Electron-Hole Plasma. *Nano Lett.* **2015**, *15* (9), 6187–6192.

(101) Shcherbakov, M. R.; Vabishchevich, P. P.; Shorokhov, A. S.; Chong, K. E.; Choi, D. Y.; Staude, I.; Miroshnichenko, A. E.; Neshev, D. N.; Fedyanin, A. A.; Kivshar, Y. S. Ultrafast All-Optical Switching with Magnetic Resonances in Nonlinear Dielectric Nanostructures. *Nano Lett.* **2015**, *15* (10), 6985–6990.

(102) Baranov, D. G.; Makarov, S. V.; Krasnok, A. E.; Belov, P. A.; Alù, A. Tuning of near- and Far-Field Properties of All-Dielectric Dimer Nanoantennas via Ultrafast Electron-Hole Plasma Photoexcitation. *Laser Photon. Rev.* **2016**, *10* (6), 1009–1015.

(103) Baranov, D. G.; Makarov, S. V.; Milichko, V. A.; Kudryashov, S. I.; Krasnok, A. E.; Belov, P. A. Nonlinear Transient Dynamics of Photoexcited Resonant Silicon

Nanostructures. *ACS Photonics* **2016**, *3* (9), 1546–1551.

(104) Savelev, R. S.; Sergaeva, O. N.; Baranov, D. G.; Krasnok, A. E.; Alù, A. Dynamically Reconfigurable Metal-Semiconductor Yagi-Uda Nanoantenna. *Phys. Rev. B* **2017**, *95* (23), 235409.

(105) Della Valle, G.; Hopkins, B.; Ganzer, L.; Stoll, T.; Rahmani, M.; Longhi, S.; Kivshar, Y. S.; De Angelis, C.; Neshev, D. N.; Cerullo, G. Nonlinear Anisotropic Dielectric Metasurfaces for Ultrafast Nanophotonics. *ACS Photonics* **2017**, 1–20.

(106) Savelev, R. S.; Filonov, D. S.; Kapitanova, P. V.; Krasnok, A. E.; Miroshnichenko, A. E.; Belov, P. A.; Kivshar, Y. S. Bending of Electromagnetic Waves in All-Dielectric Particle Array Waveguides. *Appl. Phys. Lett.* **2014**, *105* (18), 181116.

(107) Li, S. V.; Baranov, D. G.; Krasnok, A. E.; Belov, P. A. All-Dielectric Nanoantennas for Unidirectional Excitation of Electromagnetic Guided Modes. *Appl. Phys. Lett.* **2015**, *107* (17).

(108) Savelev, R. S.; Filonov, D. S.; Petrov, M. I.; Krasnok, A. E.; Belov, P. A.; Kivshar, Y. S. Resonant Transmission of Light in Chains of High-Index Dielectric Particles. *Phys. Rev. B - Condens. Matter Mater. Phys.* **2015**, *92* (15), 155415.

(109) Savelev, R. S.; Yulin, A. V.; Krasnok, A. E.; Kivshar, Y. S. Solitary Waves in Chains of High-Index Dielectric Nanoparticles. *ACS Photonics* **2016**, *3* (10), 1869–1876.

(110) Lalanne, P.; Morris, G. M. Antireflection Behavior of Silicon Subwavelength Periodic Structures for Visible Light. *Nanotechnology* **1999**, *8* (2), 53–56.

(111) Spinelli, P.; Verschuuren, M. A.; Polman, A. Broadband Omnidirectional Antireflection Coating Based on Subwavelength Surface Mie Resonators. *Nat. Commun.* **2012**, *3*, 692.

(112) Baryshnikova, K. V.; Petrov, M. I.; Babicheva, V. E.; Belov, P. A. Plasmonic and Silicon Spherical Nanoparticle Antireflective Coatings. *Sci. Rep.* **2016**, *6* (1), 22136.

(113) Proust, J.; Fehrembach, A.-L.; Bedu, F.; Ozerov, I.; Bonod, N. Optimized 2D Array of

Thin Silicon Pillars for Efficient Antireflective Coatings in the Visible Spectrum. *Sci. Rep.* **2016**, *6* (1), 24947.

(114) Odebo Länk, N.; Verre, R.; Johansson, P.; Käll, M. Large-Scale Silicon Nanophotonic Metasurfaces with Polarization Independent Near-Perfect Absorption. *Nano Lett.* **2017**, *17* (5), 3054–3060.

(115) Proust, J.; Bedu, F.; Gallas, B.; Ozerov, I.; Bonod, N. All-Dielectric Colored Metasurfaces with Silicon Mie Resonators. *ACS Nano* **2016**, *10* (8), 7761–7767.

(116) Sun, S.; Zhou, Z.; Zhang, C.; Gao, Y.; Duan, Z.; Xiao, S.; Song, Q. All-Dielectric Full-Color Printing with TiO2 Metasurfaces. *ACS Nano* **2017**, *11* (5), 4445–4452.

(117) Zhu, X.; Yan, W.; Levy, U.; Mortensen, N. A.; Kristensen, A. Resonant Laser Printing of Structural Colors on High-Index Dielectric Metasurfaces. *Sci. Adv.* **2017**, *3* (5), e1602487.

(118) Flauraud, V.; Reyes, M.; Paniagua-Domínguez, R.; Kuznetsov, A. I.; Brugger, J. Silicon Nanostructures for Bright Field Full Color Prints. *ACS Photonics* **2017**, *4* (8), 1913–1919.

(119) Huang, Z.; Wang, J.; Liu, Z.; Xu, G.; Fan, Y.; Zhong, H.; Cao, B.; Wang, C.; Xu, K. Strong-Field-Enhanced Spectroscopy in Silicon Nanoparticle Electric and Magnetic Dipole Resonance near a Metal Surface. *J. Phys. Chem. C* **2015**, *119* (50), 28127–28135.

(120) Mannino, G.; Alberti, A.; Ruggeri, R.; Libertino, S.; Pennisi, A. R.; Faraci, G. Octahedral Faceted Si Nanoparticles as Optical Traps with Enormous Yield Amplification. *Sci. Rep.* **2015**, *5*, 8354.

(121) Dmitriev, P. A.; Baranov, D. G.; Milichko, V. A.; Makarov, S. V.; Mukhin, I. S.; Samusev, A. K.; Krasnok, A. E.; Belov, P. A.; Kivshar, Y. S. Resonant Raman Scattering from Silicon Nanoparticles Enhanced by Magnetic Response. *Nanoscale* **2016**, *8* (18), 9721–9726.

(122) Frizyuk, K.; Hasan, M.; Krasnok, A.; Alu, A.; Petrov, M. Enhancement of Inherent Raman Scattering in Dielectric Nanostructures with Electric and Magnetic Mie Resonances. **2017**, 1–9.

(123) Filonov, D. S.; Slobozhanyuk, A. P.; Krasnok, A. E.; Belov, P. A.; Nenasheva, E. A.; Hopkins, B.; Miroshnichenko, A. E.; Kivshar, Y. S. Near-Field Mapping of Fano Resonances in All-Dielectric Oligomers. *Appl. Phys. Lett.* **2014**, *104* (2), 1–4.

(124) Hopkins, B.; Filonov, D. S.; Miroshnichenko, A. E.; Monticone, F.; Alù, A.; Kivshar, Y. S. Interplay of Magnetic Responses in All-Dielectric Oligomers to Realize Magnetic Fano Resonances. *ACS Photonics* **2015**, *2* (6), 724–729.

(125) Yan, J.; Liu, P.; Lin, Z.; Wang, H.; Chen, H.; Wang, C.; Yang, G. Directional Fano Resonance in a Silicon Nanosphere Dimer. *ACS Nano* **2015**, *9* (3), 2968–2980.

(126) Lepeshov, S.; Krasnok, A.; Mukhin, I.; Zuev, D.; Gudovskikh, A.; Milichko, V.; Belov, P.; Miroshnichenko, A. Fine-Tuning of the Magnetic Fano Resonance in Hybrid Oligomers via Fs-Laser-Induced Reshaping. *ACS Photonics* **2017**, *4* (3), 536–543.

(127) Abujetas, D. R.; Mandujano, M. A. G.; Méndez, E. R.; Sánchez-Gil, J. A. High-Contrast Fano Resonances in Single Semiconductor Nanorods. *ACS Photonics* **2017**, *4* (7), 1814–1821.

(128) Krasnok, A. E.; Slobozhanyuk, A. P.; Simovski, C. R.; Tretyakov, S. A.; Poddubny, A. N.; Miroshnichenko, A. E.; Kivshar, Y. S.; Belov, P. A. Antenna Model of the Purcell Effect. *Sci. Rep.* **2015**, *5* (1), 12956.

(129) Krasnok, A.; Glybovski, S.; Petrov, M.; Makarov, S.; Savelev, R.; Belov, P.; Simovski, C.; Kivshar, Y. Demonstration of the Enhanced Purcell Factor in All-Dielectric Structures. *Appl. Phys. Lett.* **2016**, *108* (21), 211105.

(130) Baranov, D. G.; Savelev, R. S.; Li, S. V.; Krasnok, A. E.; Alù, A. Modifying Magnetic Dipole Spontaneous Emission with Nanophotonic Structures. *Laser Photonics Rev.* **2017**, *11* (3), 1–16.

(131) Li, J.; Verellen, N.; Van Dorpe, P. Enhancing Magnetic Dipole Emission by a Nano-Doughnut-Shaped Silicon Disk. *ACS Photonics* **2017**, *4* (8), 1893–1898.

(132) Yang, Y.; Miller, O. D.; Christensen, T.; Joannopoulos, J. D.; Soljačić, M. Low-Loss Plasmonic Dielectric Nanoresonators. *Nano Lett.* **2017**, *17* (5), 3238–3245.

(133) Wang, H.; Ke, Y.; Xu, N.; Zhan, R.; Zheng, Z.; Wen, J.; Yan, J.; Liu, P.; Chen, J.; She, J.; Zhang, Y.; Liu, F.; Chen, H.; Deng, S. Resonance Coupling in Silicon Nanosphere-J-Aggregate Heterostructures. *Nano Lett.* **2016**, *16* (11), 6886–6895.

(134) Mie, G. Beiträge Zur Optik Trüber Medien, Speziell Kolloidaler Metallösungen. *Ann. Phys.* **1908**, *330* (3), 377–445.

(135) Bohren, Craig F, Huffman, D. R. *Absorption and Scattering of Light by Small Particles*; Bohren, C. F., Huffman, D. R., Eds.; Wiley-VCH Verlag GmbH: Weinheim, Germany, Germany, 1998.

(136) Luk'yanchuk, B.; Zheludev, N. I.; Maier, S. A.; Halas, N. J.; Nordlander, P.; Giessen, H.; Chong, C. T. The Fano Resonance in Plasmonic Nanostructures and Metamaterials. *Nat. Mater.* **2010**, *9* (9), 707–715.

(137) Alù, A.; Engheta, N. Polarizabilities and Effective Parameters for Collections of Spherical Nanoparticles Formed by Pairs of Concentric Double-Negative, Single-Negative, And∕or Double-Positive Metamaterial Layers. *J. Appl. Phys.* **2005**, *97* (9), 94310.

(138) Sauvan, C.; Hugonin, J. P.; Maksymov, I. S.; Lalanne, P. Theory of the Spontaneous Optical Emission of Nanosize Photonic and Plasmon Resonators. *Phys. Rev. Lett.* **2013**, *110* (23), 1–5.

(139) Bliokh, K. Y.; Bliokh, Y. P.; Freilikher, V.; Savel'ev, S.; Nori, F. Colloquium: Unusual Resonators: Plasmonics, Metamaterials, and Random Media. *Rev. Mod. Phys.* **2008**, *80* (4), 1201–1213.

(140) Johnson, P. B.; Christy, R. W. Optical Constants of the Noble Metals. *Phys. Rev. B*
45

**1972**, *6* (12), 4370–4379.

(141) Vuye, G.; Fisson, S.; Nguyen Van, V.; Wang, Y.; Rivory, J.; Abelès, F. Temperature Dependence of the Dielectric Function of Silicon Using in Situ Spectroscopic Ellipsometry. *Thin Solid Films* **1993**, *233* (1–2), 166–170.

(142) Alù, A.; Engheta, N. The Quest for Magnetic Plasmons at Optical Frequencies. *Opt. Express* **2009**, *17* (7), 5723–5730.

(143) Monticone, F.; Alù, A. The Quest for Optical Magnetism: From Split-Ring Resonators to Plasmonic Nanoparticles and Nanoclusters. *J. Mater. Chem. C* **2014**, *2* (43), 9059–9072.

(144) Lindquist, N. C.; Nagpal, P.; McPeak, K. M.; Norris, D. J.; Oh, S.-H. Engineering Metallic Nanostructures for Plasmonics and Nanophotonics. *Reports Prog. Phys.* **2012**, *75* (3), 36501.

(145) Maksymov, I. S.; Staude, I.; Miroshnichenko, A. E.; Kivshar, Y. S. Optical Yagi-Uda Nanoantennas. *Nanophotonics* **2012**, *1* (1), 65–81.

(146) Crut, A.; Maioli, P.; Del Fatti, N.; Vallée, F. Optical Absorption and Scattering Spectroscopies of Single Nano-Objects. *Chem. Soc. Rev.* **2014**, *43* (11), 3921.

(147) Baffou, G.; Quidant, R. Thermo-Plasmonics: Using Metallic Nanostructures as Nano-Sources of Heat. *Laser Photon. Rev.* **2013**, *7* (2), 171–187.

(148) Rahmani, M.; Xu, L.; Miroshnichenko, A. E.; Komar, A.; Camacho-Morales, R.; Chen, H.; Zárate, Y.; Kruk, S.; Zhang, G.; Neshev, D. N.; Kivshar, Y. S. Reversible Thermal Tuning of All-Dielectric Metasurfaces. *Adv. Funct. Mater.* **2017**, *27* (31), 1700580.

(149) Khurgin, J. B.; Sun, G.; Chen, W. T.; Tsai, W.-Y.; Tsai, D. P. Ultrafast Thermal Nonlinearity. *Sci. Rep.* **2016**, *5* (1), 17899.

(150) Baffou, G.; Quidant, R.; Girard, C. Thermoplasmonics Modeling: A Green's Function Approach. *Phys. Rev. B* **2010**, *82* (16), 165424.

(151) Shalabney, A.; Abdulhalim, I. Sensitivity-Enhancement Methods for Surface Plasmon

Sensors. *Laser Photonics Rev.* **2011**, *5* (4), 571–606.

(152) Couture, M.; Brulé, T.; Laing, S.; Cui, W.; Sarkar, M.; Charron, B.; Faulds, K.; Peng, W.; Canva, M.; Masson, J.-F. High Figure of Merit (FOM) of Bragg Modes in Au-Coated Nanodisk Arrays for Plasmonic Sensing. *Small* **2017**, 1700908.

(153) Jeong, H.-H.; Mark, A. G.; Alarcón-Correa, M.; Kim, I.; Oswald, P.; Lee, T.-C.; Fischer, P. Dispersion and Shape Engineered Plasmonic Nanosensors. *Nat. Commun.* **2016**, *7*, 11331.

(154) Taylor, A. B.; Zijlstra, P. Single-Molecule Plasmon Sensing: Current Status and Future Prospects. *ACS Sensors* **2017**, *2* (8), 1103–1122.

(155) Jackman, J. A.; Rahim Ferhan, A.; Cho, N.-J. Nanoplasmonic Sensors for Biointerfacial Science. *Chem. Soc. Rev.* **2017**, *46* (12), 3615–3660.

(156) Zhao, Y.; Engheta, N.; Alù, A. Effects of Shape and Loading of Optical Nanoantennas on Their Sensitivity and Radiation Properties. *J. Opt. Soc. Am. B* **2011**, *28* (5), 1266.

(157) Bontempi, N.; Chong, K. E.; Orton, H. W.; Staude, I.; Choi, D.-Y.; Alessandri, I.; Kivshar, Y. S.; Neshev, D. N. Highly Sensitive Biosensors Based on All-Dielectric Nanoresonators. *Nanoscale* **2017**, *9* (15), 4972–4980.

(158) Lodahl, P.; Mahmoodian, S.; Stobbe, S. Interfacing Single Photons and Single Quantum Dots with Photonic Nanostructures. *Rev. Mod. Phys.* **2015**, *87* (2), 347–400.

(159) Fano, U. Effects of Configuration Interaction on Intensities and Phase Shifts. *Phys. Rev.* **1961**, *124* (6), 1866–1878.

(160) Mirin, N. A.; Bao, K.; Nordlander, P. Fano Resonances in Plasmonic Nanoparticle Aggregates. *J. Phys. Chem. A* **2009**, *113* (16), 4028–4034.

(161) Miroshnichenko, A. E.; Flach, S.; Kivshar, Y. S. Fano Resonances in Nanoscale Structures. *Rev. Mod. Phys.* **2010**, *82* (3), 2257–2298.

(162) Wu, C.; Arju, N.; Kelp, G.; Fan, J. a; Dominguez, J.; Gonzales, E.; Tutuc, E.; Brener, I.; Shvets, G. Spectrally Selective Chiral Silicon Metasurfaces Based on Infrared Fano

Resonances. *Nat. Commun.* **2014**, *5* (May), 3892.

(163) Qin, F.; Lai, Y.; Yang, J.; Cui, X.; Ma, H.; Wang, J.; Lin, H.-Q. Deep Fano Resonance with Strong Polarization Dependence in Gold Nanoplate–nanosphere Heterodimers. *Nanoscale* **2017**.

(164) Limonov, M. F.; Rybin, M. V.; Poddubny, A. N.; Kivshar, Y. S. Fano Resonances in Photonics. *Nat. Photonics* **2017**, *11* (9), 543–554.

(165) Kroner, M.; Govorov, A. O.; Remi, S.; Biedermann, B.; Seidl, S.; Badolato, A.; Petroff, P. M.; Zhang, W.; Barbour, R.; Gerardot, B. D.; Warburton, R. J.; Karrai, K. The Nonlinear Fano Effect. *Nature* **2008**, *451* (7181), 1022–1022.

(166) Ridolfo, A.; Di Stefano, O.; Fina, N.; Saija, R.; Savasta, S. Quantum Plasmonics with Quantum Dot-Metal Nanoparticle Molecules: Influence of the Fano Effect on Photon Statistics. *Phys. Rev. Lett.* **2010**, *105* (26), 263601.

(167) Hartsfield, T.; Chang, W.-S.; Yang, S.-C.; Ma, T.; Shi, J.; Sun, L.; Shvets, G.; Link, S.; Li, X. Single Quantum Dot Controls a Plasmonic Cavity's Scattering and Anisotropy. *Proc. Natl. Acad. Sci.* **2015**, *112* (40), 12288–12292.

(168) Shafiei, F.; Monticone, F.; Le, K. Q.; Liu, X.-X.; Hartsfield, T.; Alù, A.; Li, X. A Subwavelength Plasmonic Metamolecule Exhibiting Magnetic-Based Optical Fano Resonance. *Nat. Nanotechnol.* **2013**, *8* (2), 95–99.

(169) Ming, T.; Chen, H.; Jiang, R.; Li, Q.; Wang, J. Plasmon-Controlled Fluorescence: Beyond the Intensity Enhancement. *J. Phys. Chem. Lett.* **2012**, *3* (2), 191–202.

(170) Forster, T. Energiewanderung Und Fluoreszenz. *Naturwissenschaften* **1946**, *33* (6), 166–175.

(171) Krainer, G.; Hartmann, A.; Schlierf, M. FarFRET: Extending the Range in Single-Molecule FRET Experiments beyond 10 Nm. *Nano Lett.* **2015**, *15* (9), 5826–5829.

(172) De Torres, J.; Mivelle, M.; Moparthi, S. B.; Rigneault, H.; Van Hulst, N. F.; García-Parajó, M. F.; Margeat, E.; Wenger, J. Plasmonic Nanoantennas Enable Forbidden

Förster Dipole-Dipole Energy Transfer and Enhance the FRET Efficiency. *Nano Lett.* **2016**, *16* (10), 6222–6230.

(173) Wang, M.; Bangalore Rajeeva, B.; Scarabelli, L.; Perillo, E. P.; Dunn, A. K.; Liz-Marzán, L. M.; Zheng, Y. Molecular-Fluorescence Enhancement via Blue-Shifted Plasmon-Induced Resonance Energy Transfer. *J. Phys. Chem. C* **2016**, *120* (27), 14820–14827.

(174) Wang, M.; Hartmann, G.; Wu, Z.; Scarabelli, L.; Rajeeva, B. B.; Jarrett, J. W.; Perillo, E. P.; Dunn, A. K.; Liz-Marzán, L. M.; Hwang, G. S.; Zheng, Y. Controlling Plasmon-Enhanced Fluorescence via Intersystem Crossing in Photoswitchable Molecules. *Small* **2017**, 1701763.

(175) Moroz, P.; Razgoniaeva, N.; Vore, A.; Eckard, H.; Kholmicheva, N.; McDarby, A.; Razgoniaev, A. O.; Ostrowski, A. D.; Khon, D.; Zamkov, M. Plasmon Induced Energy Transfer: When the Game Is Worth the Candle. *ACS Photonics* **2017**, acsphotonics.7b00527.

(176) Xiao, M.; Jiang, R.; Wang, F.; Fang, C.; Wang, J.; Yu, J. C. Plasmon-Enhanced Chemical Reactions. *J. Mater. Chem. A* **2013**, *1* (19), 5790.

(177) Holzmeister, P.; Acuna, G. P.; Grohmann, D.; Tinnefeld, P. Breaking the Concentration Limit of Optical Single-Molecule Detection. *Chem. Soc. Rev.* **2014**, *43* (4), 1014–1028.

(178) Albella, P.; Poyli, M. A.; Schmidt, M. K.; Maier, S. A.; Moreno, F.; Sáenz, J. J.; Aizpurua, J. Low-Loss Electric and Magnetic Field-Enhanced Spectroscopy with Subwavelength Silicon Dimers. *J. Phys. Chem. C* **2013**, *117* (26), 13573–13584.

(179) Rodriguez, I.; Shi, L.; Lu, X.; Korgel, B. A.; Alvarez-Puebla, R. A.; Meseguer, F. Silicon Nanoparticles as Raman Scattering Enhancers. *Nanoscale* **2014**, *6* (11), 5666–5670.

(180) Törmä, P.; Barnes, W. L. Strong Coupling between Surface Plasmon Polaritons and

Emitters. *Reports Prog. Phys.* **2014**, *78* (1), 13901.

(181) Zeng, P.; Cadusch, J.; Chakraborty, D.; Smith, T. A.; Roberts, A.; Sader, J. E.; Davis, T. J.; Gómez, D. E. Photoinduced Electron Transfer in the Strong Coupling Regime: Waveguide–Plasmon Polaritons. *Nano Lett.* **2016**, *16* (4), 2651–2656.

(182) Kasprzak, J.; Richard, M.; Kundermann, S.; Baas, A.; Jeambrun, P.; Keeling, J. M. J.; Marchetti, F. M.; Szymańska, M. H.; André, R.; Staehli, J. L.; Savona, V.; Littlewood, P. B.; Deveaud, B.; Dang, L. S. Bose–Einstein Condensation of Exciton Polaritons. *Nature* **2006**, *443* (7110), 409–414.

(183) Scafirimuto, F.; Urbonas, D.; Scherf, U.; Mahrt, R. F.; Stöferle, T. Room-Temperature Exciton-Polariton Condensation in a Tunable Zero-Dimensional Microcavity. *ACS Photonics* **2017**, acsphotonics.7b00557.

(184) Hoi, I. C.; Wilson, C. M.; Johansson, G.; Palomaki, T.; Peropadre, B.; Delsing, P. Demonstration of a Single-Photon Router in the Microwave Regime. *Phys. Rev. Lett.* **2011**, *107* (7), 73601.

(185) Volz, T.; Reinhard, A.; Winger, M.; Badolato, A.; Hennessy, K. J.; Hu, E. L.; Imamoğlu, A. Ultrafast All-Optical Switching by Single Photons. *Nat. Photonics* **2012**, *6* (9), 607–611.

(186) Sanvitto, D.; Kéna-Cohen, S. The Road towards Polaritonic Devices. *Nat. Mater.* **2016**, *15* (10), 1061–1073.

(187) Liew, T. C. H.; Kavokin, A. V.; Shelykh, I. A. Optical Circuits Based on Polariton Neurons in Semiconductor Microcavities. *Phys. Rev. Lett.* **2008**, *101* (1), 16402.

(188) Hutchison, J. A.; Schwartz, T.; Genet, C.; Devaux, E.; Ebbesen, T. W. Modifying Chemical Landscapes by Coupling to Vacuum Fields. *Angew. Chemie - Int. Ed.* **2012**, *51* (7), 1592–1596.

(189) Feist, J.; Galego, J.; García-Vidal, F. J. Polaritonic Chemistry with Organic Molecules. *ACS Photonics* **2017**, acsphotonics.7b00680.

(190) Todisco, F.; De Giorgi, M.; Esposito, M.; De Marco, L.; Zizzari, A.; Bianco, M.; Dominici, L.; Ballarini, D.; Arima, V.; Gigli, G.; Sanvitto, D. Ultrastrong Plasmon-Exciton Coupling by Dynamic Molecular Aggregation. *ACS Photonics* **2017**, acsphotonics.7b00554.

(191) Choi, H.; Heuck, M.; Englund, D. Self-Similar Nanocavity Design with Ultrasmall Mode Volume for Single-Photon Nonlinearities. *Phys. Rev. Lett.* **2017**, *118* (22), 223605.

(192) Koenderink, A. F. On the Use of Purcell Factors for Plasmon Antennas. *Opt. Lett.* **2010**, *35* (24), 4208.

(193) Derom, S.; Vincent, R.; Bouhelier, A.; Colas des Francs, G. Resonance Quality, Radiative/ohmic Losses and Modal Volume of Mie Plasmons. *EPL (Europhysics Lett.* **2012**, *98* (4), 47008.

(194) Kristensen, P. T.; Hughes, S. Modes and Mode Volumes of Leaky Optical Cavities and Plasmonic Nanoresonators. *ACS Photonics* **2014**, *1* (1), 2–10.

(195) Muljarov, E. A.; Langbein, W. Exact Mode Volume and Purcell Factor of Open Optical Systems. *Phys. Rev. B* **2016**, *94* (23), 235438.

(196) Zambrana-Puyalto, X.; Bonod, N. Purcell Factor of Spherical Mie Resonators. *Phys. Rev. B* **2015**, *91* (19), 195422.

(197) Shahbazyan, T. V. Mode Volume, Energy Transfer, and Spaser Threshold in Plasmonic Systems with Gain. *ACS Photonics* **2017**, *4* (4), 1003–1008.

(198) Wu, X.; Gray, S. K.; Pelton, M. Quantum-Dot-Induced Transparency in a Nanoscale Plasmonic Resonator. *Opt. Express* **2010**, *18* (23), 23633.

(199) Antosiewicz, T. J.; Apell, S. P.; Shegai, T. Plasmon-Exciton Interactions in a Core-Shell Geometry: From Enhanced Absorption to Strong Coupling. *ACS Photonics* **2014**, *1* (5), 454–463.

(200) Zengin, G.; Johansson, G.; Johansson, P.; Antosiewicz, T. J.; Käll, M.; Shegai, T.

Approaching the Strong Coupling Limit in Single Plasmonic Nanorods Interacting with J-Aggregates. *Sci. Rep.* **2013**, *3* (1), 3074.

(201) Abid, I.; Chen, W.; Yuan, J.; Bohloul, A.; Najmaei, S.; Avendano, C.; Péchou, R.; Mlayah, A.; Lou, J. Temperature-Dependent Plasmon-Exciton Interactions in Hybrid Au/MoSe2 Nanostructures. *ACS Photonics* **2017**, *4* (7), 1653–1660.

(202) Hensen, M.; Heilpern, T.; Gray, S. K.; Pfeiffer, W. Strong Coupling and Entanglement of Quantum Emitters Embedded in a Nanoantenna Enhanced Plasmonic Cavity. *ACS Photonics* **2017**, acsphotonics.7b00717.

(203) Stephens, W. E. Proceedings of the American Physical Society. *Phys. Rev.* **1946**, *69* (5–6), 246–260.

(204) Barthes, J.; Colas Des Francs, G.; Bouhelier, A.; Weeber, J. C.; Dereux, A. Purcell Factor for a Point-like Dipolar Emitter Coupled to a Two-Dimensional Plasmonic Waveguide. *Phys. Rev. B - Condens. Matter Mater. Phys.* **2011**, *84* (7), 73403.

(205) Poddubny, A. N.; Belov, P. A.; Ginzburg, P.; Zayats, A. V.; Kivshar, Y. S. Microscopic Model of Purcell Enhancement in Hyperbolic Metamaterials. *Phys. Rev. B - Condens. Matter Mater. Phys.* **2012**, *86* (3), 35148.

(206) Beams, R.; Smith, D.; Johnson, T. W.; Oh, S. H.; Novotny, L.; Vamivakas, A. N. Nanoscale Fluorescence Lifetime Imaging of an Optical Antenna with a Single Diamond NV Center. *Nano Lett.* **2013**, *13* (8), 3807–3811.

(207) Aigouy, L.; Cazé, A.; Gredin, P.; Mortier, M.; Carminati, R. Mapping and Quantifying Electric and Magnetic Dipole Luminescence at the Nanoscale. *Phys. Rev. Lett.* **2014**, *113* (7), 76101.

(208) Bogdanov, S.; Shalaginov, M. Y.; Akimov, A.; Lagutchev, A. S.; Kapitanova, P.; Liu, J.; Woods, D.; Ferrera, M.; Belov, P.; Irudayaraj, J.; Boltasseva, A.; Shalaev, V. M. Electron Spin Contrast of Purcell-Enhanced Nitrogen-Vacancy Ensembles in Nanodiamonds. *Phys. Rev. B* **2017**, *96* (3), 35146.

(209) Lončar, M.; Scherer, A.; Qiu, Y. *Photonic Crystal Laser Sources for Chemical Detection*; 2003; Vol. 82.

(210) Devilez, A.; Stout, B.; Bonod, N. Compact Metallo-Dielectric Optical Antenna for Ultra Directional and Enhanced Radiative Emission. *ACS Nano* **2010**, *4* (6), 3390–3396.

(211) Gérard, D.; Wenger, J.; Devilez, A.; Gachet, D.; Stout, B.; Bonod, N.; Popov, E.; Rigneault, H. Strong Electromagnetic Confinement near Dielectric Microspheres to Enhance Single-Molecule Fluorescence. *Opt. Express* **2008**, *16* (19), 15297.

(212) Gérard, D.; Devilez, A.; Aouani, H.; Stout, B.; Bonod, N.; Wenger, J.; Popov, E.; Rigneault, H. Efficient Excitation and Collection of Single-Molecule Fluorescence close to a Dielectric Microsphere. *J. Opt. Soc. Am. B* **2009**, *26* (7), 1473.

(213) Wenger, J.; Rigneault, H. Photonic Methods to Enhance Fluorescence Correlation Spectroscopy and Single Molecule Fluorescence Detection. *Int. J. Mol. Sci.* **2010**, *11* (1), 206–221.

(214) Wang, Z.; Guo, W.; Li, L.; Luk'yanchuk, B.; Khan, A.; Liu, Z.; Chen, Z.; Hong, M. Optical Virtual Imaging at 50 Nm Lateral Resolution with a White-Light Nanoscope. *Nat. Commun.* **2011**, *2*, 218.

(215) Krasnok, A. E.; Miroshnichenko, A. E.; Belov, P. A.; Kivshar, Y. S. Huygens Optical Elements and Yagi—Uda Nanoantennas Based on Dielectric Nanoparticles. *JETP Lett.* **2011**, *94* (8), 593–598.

(216) Fu, Y. H.; Kuznetsov, A. I.; Miroshnichenko, A. E.; Yu, Y. F.; Luk'yanchuk, B. Directional Visible Light Scattering by Silicon Nanoparticles. *Nat. Commun.* **2013**, *4*, 1527.

(217) Tribelsky, M. I.; Geffrin, J.-M.; Litman, A.; Eyraud, C.; Moreno, F. Small Dielectric Spheres with High Refractive Index as New Multifunctional Elements for Optical Devices. *Sci. Rep.* **2015**, *5* (1), 12288.

(218) Fan, X.; Zheng, W.; Singh, D. J. Light Scattering and Surface Plasmons on Small

Spherical Particles. *Light Sci. Appl.* **2014**, *3* (6), e179.

(219) Alaee, R.; Filter, R.; Lehr, D.; Lederer, F.; Rockstuhl, C. A Generalized Kerker Condition for Highly Directive Nanoantennas. *Opt. Lett.* **2015**, *40* (11), 2645.

(220) Yao, K.; Liu, Y. Controlling Electric and Magnetic Resonances for Ultracompact Nanoantennas with Tunable Directionality. *ACS Photonics* **2016**, *3* (6), 953–963.

(221) Terekhov, P. D.; Baryshnikova, K. V.; Artemyev, Y. A.; Karabchevsky, A.; Shalin, A. S.; Evlyukhin, A. B. Multipolar Response of Nonspherical Silicon Nanoparticles in the Visible and near-Infrared Spectral Ranges. *Phys. Rev. B* **2017**, *96* (3), 35443.

(222) Cho, Y.; Huh, J.-H.; Park, K. J.; Kim, K.; Lee, J.; Lee, S. Using Highly Uniform and Smooth Selenium Colloids as Low-Loss Magnetodielectric Building Blocks of Optical Metafluids. *Opt. Express* **2017**, *25* (12), 13822.

(223) Miroshnichenko, A. E. All-Dielectric Optical Nanoantennas. In *IEEE Antennas and Propagation Society, AP-S International Symposium (Digest)*; 2015; Vol. 2015–Octob, pp 601–602.

(224) C. A. Balanis. *Antenna Theory: Analysis and Design*; New York ; Brisbane : J. Wiley, 1997.

(225) Li, J.; Salandrino, A.; Engheta, N. Shaping Light Beams in the Nanometer Scale: A Yagi-Uda Nanoantenna in the Optical Domain. *Phys. Rev. B* **2007**, *76* (24), 245403.

(226) Lobanov, S. V.; Weiss, T.; Dregely, D.; Giessen, H.; Gippius, N. A.; Tikhodeev, S. G. Emission Properties of an Oscillating Point Dipole from a Gold Yagi-Uda Nanoantenna Array. *Phys. Rev. B - Condens. Matter Mater. Phys.* **2012**, *85* (15), 155137.

(227) Li, J.; Salandrino, A.; Engheta, N. Optical Spectrometer at the Nanoscale Using Optical Yagi-Uda Nanoantennas. *Phys. Rev. B - Condens. Matter Mater. Phys.* **2009**, *79* (19), 195104.

(228) Rolly, B.; Stout, B.; Bonod, N. Boosting the Directivity of Optical Antennas with Magnetic and Electric Dipolar Resonant Particles. *Opt. Express* **2012**, *20* (18), 20376.

(229) Liu, W.; Miroshnichenko, A. E.; Neshev, D. N.; Kivshar, Y. S. Broadband Unidirectional Scattering by Magneto-Electric Core-Shell Nanoparticles. *ACS Nano* **2012**, *6* (6), 5489–5497.

(230) Rolly, B.; Bebey, B.; Bidault, S.; Stout, B.; Bonod, N. Promoting Magnetic Dipolar Transition in Trivalent Lanthanide Ions with Lossless Mie Resonances. *Phys. Rev. B - Condens. Matter Mater. Phys.* **2012**, *85* (24), 2–7.

(231) Filonov, D. S.; Krasnok, A. E.; Slobozhanyuk, A. P.; Kapitanova, P. V.; Nenasheva, E. A.; Kivshar, Y. S.; Belov, P. A. Experimental Verification of the Concept of All-Dielectric Nanoantennas. *Appl. Phys. Lett.* **2012**, *100* (20), 1–5.

(232) Liu, Y. G.; Choy, W. C. H.; Sha, W. E. I.; Chew, W. C. Unidirectional and Wavelength-Selective Photonic Sphere-Array Nanoantennas. *Opt. Lett.* **2012**, *37* (11), 2112–2114.

(233) Schmidt, M. K.; Esteban, R.; Sáenz, J. J.; Suárez-Lacalle, I.; Mackowski, S.; Aizpurua, J. Dielectric Antennas - a Suitable Platform for Controlling Magnetic Dipolar Emission: Errata. *Opt. Express* **2012**, *20* (17), 18609.

(234) Zou, L.; Withayachumnankul, W.; Shah, C. M.; Mitchell, A.; Bhaskaran, M.; Sriram, S.; Fumeaux, C. Dielectric Resonator Nanoantennas at Visible Frequencies. *Opt. Express* **2013**, *21* (1), 83–90.

(235) Liu, W.; Miroshnichenko, A. E.; Kivshar, Y. S. Control of Light Scattering by Nanoparticles with Optically-Induced Magnetic Responses. *Chinese Phys. B* **2014**, *23* (4), 47806.

(236) Staude, I.; Miroshnichenko, A. E.; Decker, M.; Fofang, N. T.; Liu, S.; Gonzales, E.; Dominguez, J.; Luk, T. S.; Neshev, D. N.; Brener, I.; Kivshar, Y. Tailoring Directional Scattering through Magnetic and Electric Resonances in Subwavelength Silicon Nanodisks. *ACS Nano* **2013**, *7* (9), 7824–7832.

(237) Powell, D. A. Interference between the Modes of an All-Dielectric Meta-Atom. *Phys.*

*Rev. Appl.* **2017**, *7* (3), 34006.

(238) Boudarham, G.; Abdeddaim, R.; Bonod, N. Enhancing the Magnetic Field Intensity with a Dielectric Gap Antenna. *Appl. Phys. Lett.* **2014**, *104* (2), 21117.

(239) Mitrofanov, O.; Dominec, F.; Kužel, P.; Reno, J. L.; Brener, I.; Chung, U.-C.; Elissalde, C.; Maglione, M.; Mounaix, P. Near-Field Probing of Mie Resonances in Single $TiO_2$ Microspheres at Terahertz Frequencies. *Opt. Express* **2014**, *22* (19), 23034.

(240) Bakker, R. M.; Permyakov, D.; Yu, Y. F.; Markovich, D.; Paniagua-Domínguez, R.; Gonzaga, L.; Samusev, A.; Kivshar, Y.; Luk'yanchuk, B.; Kuznetsov, A. I. Magnetic and Electric Hotspots with Silicon Nanodimers. *Nano Lett.* **2015**, *15* (3), 2137–2142.

(241) Krasnok, A. E.; Simovski, C. R.; Belov, P. A.; Kivshar, Y. S. Superdirective Dielectric Nanoantennas. *Nanoscale* **2014**, *6* (13), 7354–7361.

(242) Rolly, B.; Geffrin, J.-M.; Abdeddaim, R.; Stout, B.; Bonod, N. Controllable Emission of a Dipolar Source Coupled with a Magneto-Dielectric Resonant Subwavelength Scatterer. *Sci. Rep.* **2013**, *3* (1), 3063.

(243) Bigourdan, F.; Marquier, F.; Hugonin, J.-P.; Greffet, J.-J. Design of Highly Efficient Metallo-Dielectric Patch Antennas for Single-Photon Emission. *Opt. Express* **2014**, *22* (3), 2337.

(244) Rahmani, M.; Luk'yanchuk, B.; Hong, M. Fano Resonance in Novel Plasmonic Nanostructures. *Laser Photonics Rev.* **2013**, *7* (3), 329–349.

(245) Hentschel, M.; Saliba, M.; Vogelgesang, R.; Giessen, H.; Alivisatos, A. P.; Liu, N. Transition from Isolated to Collective Modes in Plasmonic Oligomers. *Nano Lett.* **2010**, *10* (7), 2721–2726.

(246) Deng, H.-D.; Chen, X.-Y.; Xu, Y.; Miroshnichenko, A. E. Single Protein Sensing with Asymmetric Plasmonic Hexamer via Fano Resonance Enhanced Two-Photon Luminescence. *Nanoscale* **2015**, *7* (48), 20405–20413.

(247) Verre, R.; Yang, Z. J.; Shegai, T.; Käll, M. Optical Magnetism and Plasmonic Fano

Resonances in Metal-Insulator-Metal Oligomers. *Nano Lett.* **2015**, *15* (3), 1952–1958.

(248) Jing, X.; Ye, Q.; Hong, Z.; Zhu, D.; Shi, G. Broadband Electromagnetic Dipole Scattering by Coupled Multiple Nanospheres. *Superlattices Microstruct.* **2017**.

(249) Zywietz, U.; Schmidt, M. K.; Evlyukhin, A. B.; Reinhardt, C.; Aizpurua, J.; Chichkov, B. N. Electromagnetic Resonances of Silicon Nanoparticle Dimers in the Visible. *ACS Photonics* **2015**, *2* (7), 913–920.

(250) García-Cámara, B.; Algorri, J. F.; Cuadrado, A.; Urruchi, V.; Sánchez-Pena, J. M.; Serna, R.; Vergaz, R. All-Optical Nanometric Switch Based on the Directional Scattering of Semiconductor Nanoparticles. *J. Phys. Chem. C* **2015**, *119* (33), 19558–19564.

(251) Karaveli, S.; Zia, R. Spectral Tuning by Selective Enhancement of Electric and Magnetic Dipole Emission. *Phys. Rev. Lett.* **2011**, *106* (19), 193004.

(252) Karaveli, S.; Wang, S.; Xiao, G.; Zia, R. Time-Resolved Energy-Momentum Spectroscopy of Electric and Magnetic Dipole Transitions in $Cr^{3+}$:MgO. *ACS Nano* **2013**, *7* (8), 7165–7172.

(253) Li, D.; Jiang, M.; Cueff, S.; Dodson, C. M.; Karaveli, S.; Zia, R. Quantifying and Controlling the Magnetic Dipole Contribution to Light Emission in Erbium-Doped Yttrium Oxide. *Phys. Rev. B* **2014**, *89* (16), 161409.

(254) Hussain, R.; Kruk, S. S.; Bonner, C. E.; Noginov, M. A.; Staude, I.; Kivshar, Y. S.; Noginova, N.; Neshev, D. N. Enhancing $Eu^{3+}$ Magnetic Dipole Emission by Resonant Plasmonic Nanostructures. *Opt. Lett.* **2015**, *40* (8), 1659.

(255) Rabouw, F. T.; Prins, P. T.; Norris, D. J. Europium-Doped $NaYF_4$ Nanocrystals as Probes for the Electric and Magnetic Local Density of Optical States throughout the Visible Spectral Range. *Nano Lett.* **2016**, *16* (11), 7254–7260.

(256) Miri, M.; Sadrara, M. Dimers and Trimers of Hollow Silicon Nanoparticles: Manipulating the Magnetic Hotspots. *J. Phys. Chem. C* **2017**, *121* (21), 11672–11679.

(257) Grinblat, G.; Li, Y.; Nielsen, M. P.; Oulton, R. F.; Maier, S. A. Efficient Third Harmonic Generation and Nonlinear Subwavelength Imaging at a Higher-Order Anapole Mode in a Single Germanium Nanodisk. *ACS Nano* **2017**, *11* (1), 953–960.

(258) Liu, S.; Vaskin, A.; Campione, S.; Wolf, O.; Sinclair, M. B.; Reno, J.; Keeler, G. A.; Staude, I.; Brener, I. Huygens' Metasurfaces Enabled by Magnetic Dipole Resonance Tuning in Split Dielectric Nanoresonators. *Nano Lett.* **2017**, *17* (7), 4297–4303.

(259) Chong, K. E.; Hopkins, B.; Staude, I.; Miroshnichenko, A. E.; Dominguez, J.; Decker, M.; Neshev, D. N.; Brener, I.; Kivshar, Y. S. Observation of Fano Resonances in All-Dielectric Nanoparticle Oligomers. *Small* **2014**, *10* (10), 1985–1990.

(260) Yang, Y.; Kravchenko, I. I.; Briggs, D. P.; Valentine, J. All-Dielectric Metasurface Analogue of Electromagnetically Induced Transparency. *Nat. Commun.* **2014**, *5*, 5753.

(261) Dmitriev, P. A.; Makarov, S. V.; Milichko, V. A.; Mukhin, I. S.; Gudovskikh, A. S.; Sitnikova, A. A.; Samusev, A. K.; Krasnok, A. E.; Belov, P. A. Laser Fabrication of Crystalline Silicon Nanoresonators from an Amorphous Film for Low-Loss All-Dielectric Nanophotonics. *Nanoscale* **2016**, *8* (9), 5043–5048.

(262) Forouzmand, A.; Mosallaei, H. All-Dielectric C-Shaped Nanoantennas for Light Manipulation: Tailoring Both Magnetic and Electric Resonances to the Desire. *Adv. Opt. Mater.* **2017**, *5* (14), 1700147.

(263) Butler, S. Z.; Hollen, S. M.; Cao, L.; Cui, Y.; Gupta, J. A.; Gutiérrez, H. R.; Heinz, T. F.; Hong, S. S.; Huang, J.; Ismach, A. F.; Johnston-Halperin, E.; Kuno, M.; Plashnitsa, V. V.; Robinson, R. D.; Ruoff, R. S.; Salahuddin, S.; Shan, J.; Shi, L.; Spencer, M. G.; Terrones, M.; Windl, W.; Goldberger, J. E. Progress, Challenges, and Opportunities in Two-Dimensional Materials beyond Graphene. *ACS Nano* **2013**, *7* (4), 2898–2926.

(264) Xia, F.; Wang, H.; Xiao, D.; Dubey, M.; Ramasubramaniam, A. Two-Dimensional Material Nanophotonics. *Nat. Photonics* **2014**, *8* (12), 899–907.

(265) Mak, K. F.; Shan, J. Photonics and Optoelectronics of 2D Semiconductor Transition

Metal Dichalcogenides. *Nat. Photonics* **2016**, *10* (4), 216–226.

(266) Zhao, W.; Ghorannevis, Z.; Chu, L.; Toh, M.; Kloc, C.; Tan, P. H.; Eda, G. Evolution of Electronic Structure in Atomically Thin Sheets of Ws 2 and wse2. *ACS Nano* **2013**, *7* (1), 791–797.

(267) Lopez-Sanchez, O.; Lembke, D.; Kayci, M.; Radenovic, A.; Kis, A. Ultrasensitive Photodetectors Based on Monolayer MoS2. *Nat. Nanotechnol.* **2013**, *8* (7), 497–501.

(268) Yin, Z.; Li, H.; Li, H.; Jiang, L.; Shi, Y.; Sun, Y.; Lu, G.; Zhang, Q.; Chen, X.; Zhang, H. Single-Layer MoS 2 Phototransistors. *ACS Nano* **2012**, *6* (1), 74–80.

(269) Withers, F.; Del Pozo-Zamudio, O.; Mishchenko, A.; Rooney, A. P.; Gholinia, A.; Watanabe, K.; Taniguchi, T.; Haigh, S. J.; Geim, A. K.; Tartakovskii, A. I.; Novoselov, K. S. Light-Emitting Diodes by Band-Structure Engineering in van Der Waals Heterostructures. *Nat. Mater.* **2015**, *14* (3), 301–306.

(270) Ye, Y.; Wong, Z. J.; Lu, X.; Ni, X.; Zhu, H.; Chen, X.; Wang, Y.; Zhang, X. Monolayer Excitonic Laser. *Nat. Photonics* **2015**, *9* (11), 733–737.

(271) Sergey Lepeshov, Mingsong Wang, Alex Krasnok, Oleg Kotov, Tianyi Zhang, He Liu, Taizhi Jiang, Brian Korgel, Mauricio Terrones, Yuebing Zheng, and A. A. Tunable Resonant Coupling Between a Single Si Nanoparticle and a WS2 Monolayer. *ArXiv e-prints* **2017**.

(272) Nie, S. and Emory, S.-R. Probing Single Molecules and Single Nanoparticles by Surface-Enhanced Raman Scattering. *Science (80-. ).* **1997**, *275* (February), 1102–1106.

(273) Maier, S. A. *Plasmonics: Fundamentals and Applications*; Springer US: Boston, MA, 2007; Vol. 677.

(274) Chan, J.; Fore, S.; Wachsman-Hogiu, S.; Huser, T. Raman Spectroscopy and Microscopy of Individual Cells and Cellular Components. *Laser Photon. Rev.* **2008**, *2* (5), 325–349.

(275) Rui Tan, J. M.; Ruan, J. J.; Lee, H. K.; Phang, I. Y.; Ling, X. Y. A Large-Scale

Superhydrophobic Surface-Enhanced Raman Scattering (SERS) Platform Fabricated via Capillary Force Lithography and Assembly of Ag Nanocubes for Ultratrace Molecular Sensing. *Phys. Chem. Chem. Phys.* **2014**, *16* (48), 26983–26990.

(276) Antonio, K. A.; Schultz, Z. D. Advances in Biomedical Raman Microscopy. *Anal. Chem.* **2014**, *86* (1), 30–46.

(277) Zheng, P.; Li, M.; Jurevic, R.; Cushing, S. K.; Liu, Y.; Wu, N. A Gold Nanohole Array Based Surface-Enhanced Raman Scattering Biosensor for Detection of Silver(i) and Mercury(ii) in Human Saliva. *Nanoscale* **2015**, *7* (25), 11005–11012.

(278) Alessandri, I.; Lombardi, J. R. Enhanced Raman Scattering with Dielectrics. *Chem. Rev.* **2016**, *116* (24), 14921–14981.

(279) Giannini, V.; Fernández-Domínguez, A. I.; Heck, S. C.; Maier, S. A. Plasmonic Nanoantennas: Fundamentals and Their Use in Controlling the Radiative Properties of Nanoemitters. *Chem. Rev.* **2011**, *111* (6), 3888–3912.

(280) Siraj, N.; El-Zahab, B.; Hamdan, S.; Karam, T. E.; Haber, L. H.; Li, M.; Fakayode, S. O.; Das, S.; Valle, B.; Strongin, R. M.; Patonay, G.; Sintim, H. O.; Baker, G. A.; Powe, A.; Lowry, M.; Karolin, J. O.; Geddes, C. D.; Warner, I. M. Fluorescence, Phosphorescence, and Chemiluminescence. *Anal. Chem.* **2016**, *88* (1), 170–202.

(281) Li, J.-F.; Li, C.-Y.; Aroca, R. F. Plasmon-Enhanced Fluorescence Spectroscopy. *Chem. Soc. Rev.* **2017**, *46* (13), 3962–3979.

(282) Langguth, L.; Szuba, A.; Mann, S. A.; Garnett, E. C.; Koenderink, G. H.; Koenderink, A. F. Nano-Antenna Enhanced Two-Focus Fluorescence Correlation Spectroscopy. *Sci. Rep.* **2017**, *7* (1), 5985.

(283) Luo, S.; Li, Q.; Yang, Y.; Chen, X.; Wang, W.; Qu, Y.; Qiu, M. Controlling Fluorescence Emission with Split-Ring-Resonator-Based Plasmonic Metasurfaces. *Laser Photon. Rev.* **2017**, *11* (3), 1600299.

(284) Miroshnichenko, A. E.; Kivshar, Y. S. Fano Resonances in All-Dielectric Oligomers.

*Nano Lett.* **2012**, *12* (12), 6459–6463.

(285) Habteyes, T. G.; Staude, I.; Chong, K. E.; Dominguez, J.; Decker, M.; Miroshnichenko, A.; Kivshar, Y.; Brener, I. Near-Field Mapping of Optical Modes on All-Dielectric Silicon Nanodisks. *ACS Photonics* **2014**, *1* (9), 794–798.

(286) Hopkins, B.; Poddubny, A. N.; Miroshnichenko, A. E.; Kivshar, Y. S. Revisiting the Physics of Fano Resonances for Nanoparticle Oligomers. *Phys. Rev. A - At. Mol. Opt. Phys.* **2013**, *88* (5), 1–10.

(287) Yuan, H.; Khatua, S.; Zijlstra, P.; Yorulmaz, M.; Orrit, M. Thousand-Fold Enhancement of Single-Molecule Fluorescence near a Single Gold Nanorod. *Angew. Chemie - Int. Ed.* **2013**, *52* (4), 1217–1221.

(288) Jin, Y. Engineering Plasmonic Gold Nanostructures and Metamaterials for Biosensing and Nanomedicine. *Adv. Mater.* **2012**, *24* (38), 5153–5165.

(289) Valsecchi, C.; Brolo, A. G. Periodic Metallic Nanostructures as Plasmonic Chemical Sensors. *Langmuir* **2013**, *29* (19), 5638–5649.

(290) Nair, S.; Escobedo, C.; Sabat, R. G. Crossed Surface Relief Gratings as Nanoplasmonic Biosensors. *ACS Sensors* **2017**, *2* (3), 379–385.

(291) Lal, S.; Link, S.; Halas, N. J. Nano-Optics from Sensing to Waveguiding. *Nat. Photonics* **2007**, *1* (11), 641–648.

(292) Schmidt, M. A.; Lei, D. Y.; Wondraczek, L.; Nazabal, V.; Maier, S. A. Hybrid Nanoparticle–microcavity-Based Plasmonic Nanosensors with Improved Detection Resolution and Extended Remote-Sensing Ability. *Nat. Commun.* **2012**, *3* (May), 1108.

(293) Saha, K.; Agasti, S.; Kim, C.; Li, X.; Rotello, V. Gold Nanoparticles in Chemical and Biological Sensing More 288 Views. *Chem. Rev.* **2012**, *112* (5), 2739–2779.

(294) Wang, Y.; Knoll, W.; Dostalek, J. Bacterial Pathogen Surface Plasmon Resonance Biosensor Advanced by Long Range Surface Plasmons and Magnetic Nanoparticle Assays. *Anal. Chem.* **2012**, *84* (19), 8345–8350.

(295) Homola, J. Surface Plasmon Resonance Sensors for Detection of Chemical and Biological Species. *Chem. Rev.* **2008**, *108* (2), 462–493.

(296) Teh, H. F.; Peh, W. Y. X.; Su, X.; Thomsen, J. S. Characterization of Protein-DNA Interactions Using Surface Plasmon Resonance Spectroscopy with Various Assay Schemes. *Biochemistry* **2007**, *46* (8), 2127–2135.

(297) Aćimović, S. S.; Ortega, M. A.; Sanz, V.; Berthelot, J.; Garcia-Cordero, J. L.; Renger, J.; Maerkl, S. J.; Kreuzer, M. P.; Quidant, R. LSPR Chip for Parallel, Rapid, and Sensitive Detection of Cancer Markers in Serum. *Nano Lett.* **2014**, *14* (5), 2636–2641.

(298) Berto, P.; Mohamed, M. S. A.; Rigneault, H.; Baffou, G. Time-Harmonic Optical Heating of Plasmonic Nanoparticles. *Phys. Rev. B* **2014**, *90* (3), 35439.

(299) Mahmoudi, M.; Lohse, S. E.; Murphy, C. J.; Fathizadeh, A.; Montazeri, A.; Suslick, K. S. Variation of Protein Corona Composition of Gold Nanoparticles Following Plasmonic Heating. *Nano Lett.* **2014**, *14* (1), 6–12.

(300) García-Cámara, B.; Gómez-Medina, R.; Saenz, Jose, J.; Sepulveda, B. Sensing with Magnetic Dipolar Resonances in Semiconductor Nanospheres. *Opt. Express* **2013**, *21* (20), 23007–23020.

(301) Zhang, G.; Lan, C.; Bian, H.; Gao, R.; Zhou, J. Flexible, All-Dielectric Metasurface Fabricated via Nanosphere Lithography and Its Applications in Sensing. *Opt. Express* **2017**, *25* (18), 22038.

(302) Morita, M.; Ohmi, T.; Hasegawa, E.; Kawakami, M.; Ohwada, M. Growth of Native Oxide on a Silicon Surface. *J. Appl. Phys.* **1990**, *68* (3), 1272–1281.

(303) Marinakos, S. M.; Chen, S.; Chilkoti, A. Plasmonic Detection of a Model Analyte in Serum by a Gold Nanorod Sensor. *Anal. Chem.* **2007**, *79* (14), 5278–5283.

(304) Guo, Q.; Zhu, H.; Liu, F.; Zhu, A. Y.; Reed, J. C.; Yi, F.; Cubukcu, E. Silicon-on-Glass Graphene-Functionalized Leaky Cavity Mode Nanophotonic Biosensor. *ACS Photonics* **2014**, *1* (3), 221–227.

(305) Henzie, J.; Lee, M. H.; Odom, T. W. Multiscale Patterning of Plasmonic Metamaterials. *Nat. Nanotechnol.* **2007**, *2* (9), 549–554.

(306) Yavas, O.; Svedendahl, M.; Dobosz, P.; Sanz, V.; Quidant, R. On-a-Chip Biosensing Based on All-Dielectric Nanoresonators. *Nano Lett.* **2017**, *17* (7), 4421–4426.

(307) Mayer, K. M.; Hafner, J. H. Localized Surface Plasmon Resonance Sensors. *Chem. Rev.* **2011**, *111* (6), 3828–3857.

(308) Timpu, F.; Sergeyev, A.; Hendricks, N. R.; Grange, R. Second-Harmonic Enhancement with Mie Resonances in Perovskite Nanoparticles. *ACS Photonics* **2017**, *4* (1), 76–84.

(309) Staedler, D.; Magouroux, T.; Hadji, R.; Joulaud, C.; Extermann, J.; Schwung, S.; Passemard, S.; Kasparian, C.; Clarke, G.; Gerrmann, M.; Dantec, R. Le; Mugnier, Y.; Rytz, D.; Ciepielewski, D.; Galez, C.; Gerber-Lemaire, S.; Juillerat-Jeanneret, L.; Bonacina, L.; Wolf, J. P. Harmonic Nanocrystals for Biolabeling: A Survey of Optical Properties and Biocompatibility. *ACS Nano* **2012**, *6* (3), 2542–2549.

(310) Grange, R.; Lanvin, T.; Hsieh, C.-L.; Pu, Y.; Psaltis, D. Imaging with Second-Harmonic Radiation Probes in Living Tissue. *Biomed. Opt. Express* **2011**, *2* (9), 2532.

(311) Timpu, F.; Hendricks, N. R.; Petrov, M.; Ni, S.; Renaut, C.; Wolf, H.; Isa, L.; Kivshar, Y.; Grange, R. Enhanced Second-Harmonic Generation from Sequential Capillarity-Assisted Particle Assembly of Hybrid Nanodimers. *Nano Lett.* **2017**, *17* (9), 5381–5388.

(312) Marinica, D. C.; Borisov, A. G.; Shabanov, S. V. Bound States in the Continuum in Photonics. *Phys. Rev. Lett.* **2008**, *100* (18), 1–4.

(313) Hsu, C. W.; Zhen, B.; Stone, A. D.; Joannopoulos, J. D.; Soljačić, M. Bound States in the Continuum. *Nat. Rev. Mater.* **2016**, *1* (9), 16048.

(314) Monticone, F.; Alù, A. Embedded Photonic Eigenvalues in 3D Nanostructures. *Phys. Rev. Lett.* **2014**, *112* (21), 213903.

(315) Monticone, F.; Alù, A. Bound States within the Radiation Continuum in Diffraction

Gratings and the Role of Leaky Modes. *New J. Phys.* **2017**, *19* (9), 93011.

(316) Bulgakov, E. N.; Sadreev, A. F. Bloch Bound States in the Radiation Continuum in a Periodic Array of Dielectric Rods. *Phys. Rev. A - At. Mol. Opt. Phys.* **2014**, *90* (5), 53801.

(317) Bulgakov, E. N.; Maksimov, D. N. Topological Bound States in the Continuum in Arrays of Dielectric Spheres. *Phys. Rev. Lett.* **2017**, *118* (26), 267401.

(318) Underwood, S.; Mulvaney, P. Effect of the Solution Refractive Index on the Color of Gold Colloids. *Langmuir* **1994**, *10* (10), 3427–3430.

(319) Paulo, P. M. R.; Zijlstra, P.; Orrit, M.; Garcia-Fernandez, E.; Pace, T. C. S.; Viana, A. S.; Costa, S. M. B. Tip-Specific Functionalization of Gold Nanorods for Plasmonic Biosensing: Effect of Linker Chain Length. *Langmuir* **2017**, *33* (26), 6503–6510.

(320) Burgin, J.; Liu, M.; Guyot-Sionnest, P. Dielectric Sensing with Deposited Gold Bipyramids. *J. Phys. Chem. C* **2008**, *112* (49), 19278–19282.

(321) Liu, N.; Weiss, T.; Mesch, M.; Langguth, L.; Eigenthaler, U.; Hirscher, M.; Sönnichsen, C.; Giessen, H. Planar Metamaterial Analogue of Electromagnetically Induced Transparency for Plasmonic Sensing. *Nano Lett.* **2010**, *10* (4), 1103–1107.

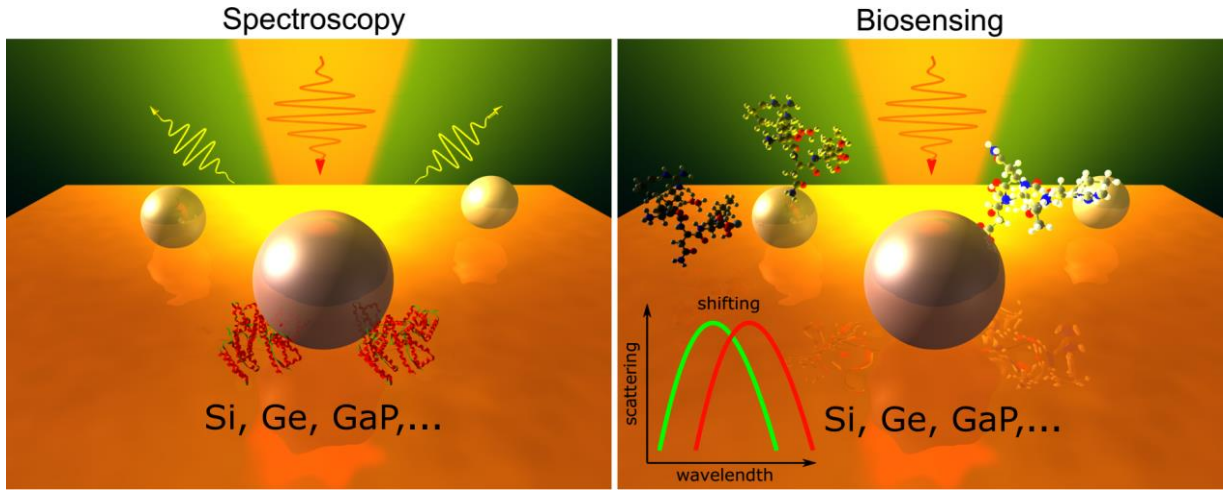

**Figure 1.** Schematic representation of topics addressed in this review: optical Spectroscopy (left) and biosensing (right) based on optically resonant dielectric nanostructures.

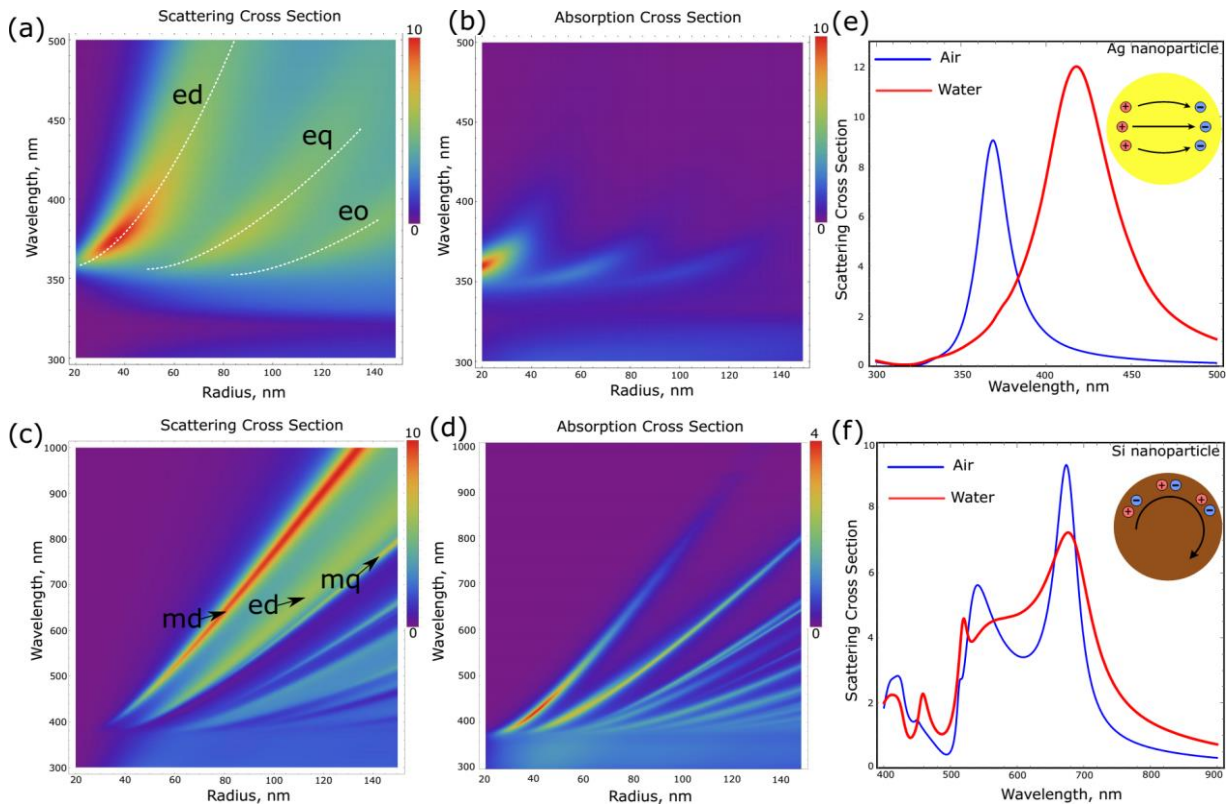

**Figure 2.** Optical properties of metallic (Ag) and dielectric (Si) nanoparticles. (a),(c) Normalized total scattering cross section ($Q_{sct}$) of the Ag nanoparticle (a) and the Si nanoparticle (c), depending on their radius (R) and wavelength. (b), (d) Normalized absorption cross section ($Q_{abs}$) of the Ag nanoparticle (b) and the Si nanoparticle (d),

depending on their radius and wavelength. (e), (f) Normalized scattering spectra of the Ag nanoparticle of radius 30 nm (e) and the Si nanoparticle (of radius 85 nm) placed in air ($\varepsilon_h = 1$, blue curves) and water ($\varepsilon_h = 1.77$, red curves). Insets: schematic diagrams representing the fundamental modes of the metallic nanoparticle (plasmon dipole moment) and the dielectric nanoparticle (magnetic dipole Mie moment).

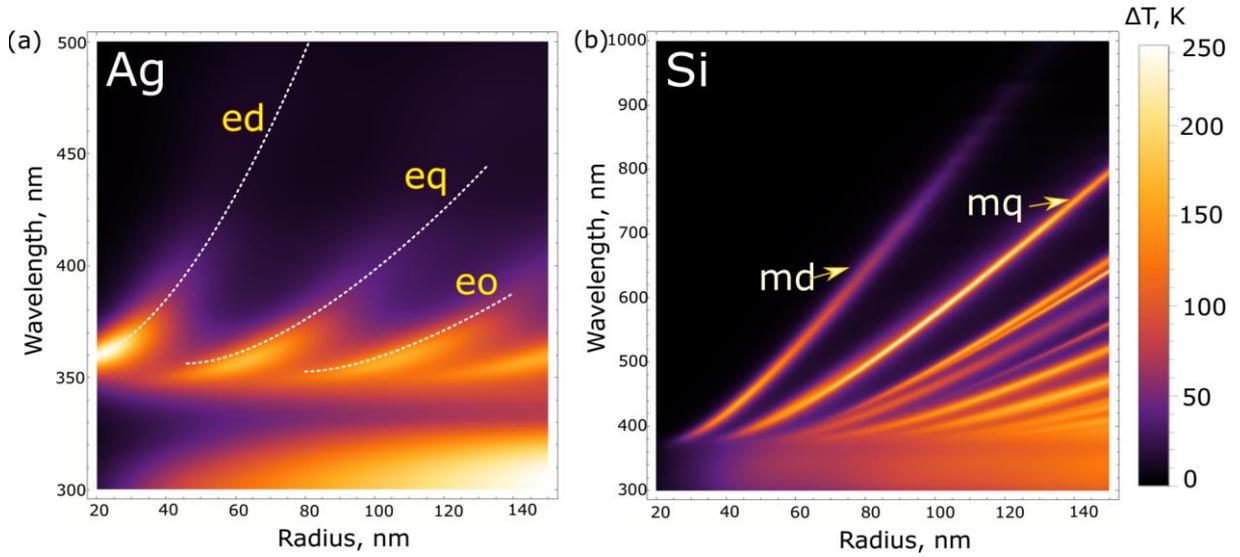

**Figure 3.** Temperature increase calculated by Eq. (6) for heating of (a) silver and (b) silicon spherical NPs in air ($\chi_2 = 0.025$ W/m K[150]) at light intensity of I=0.1 mW/μm², depending on their radius (R) and wavelength.

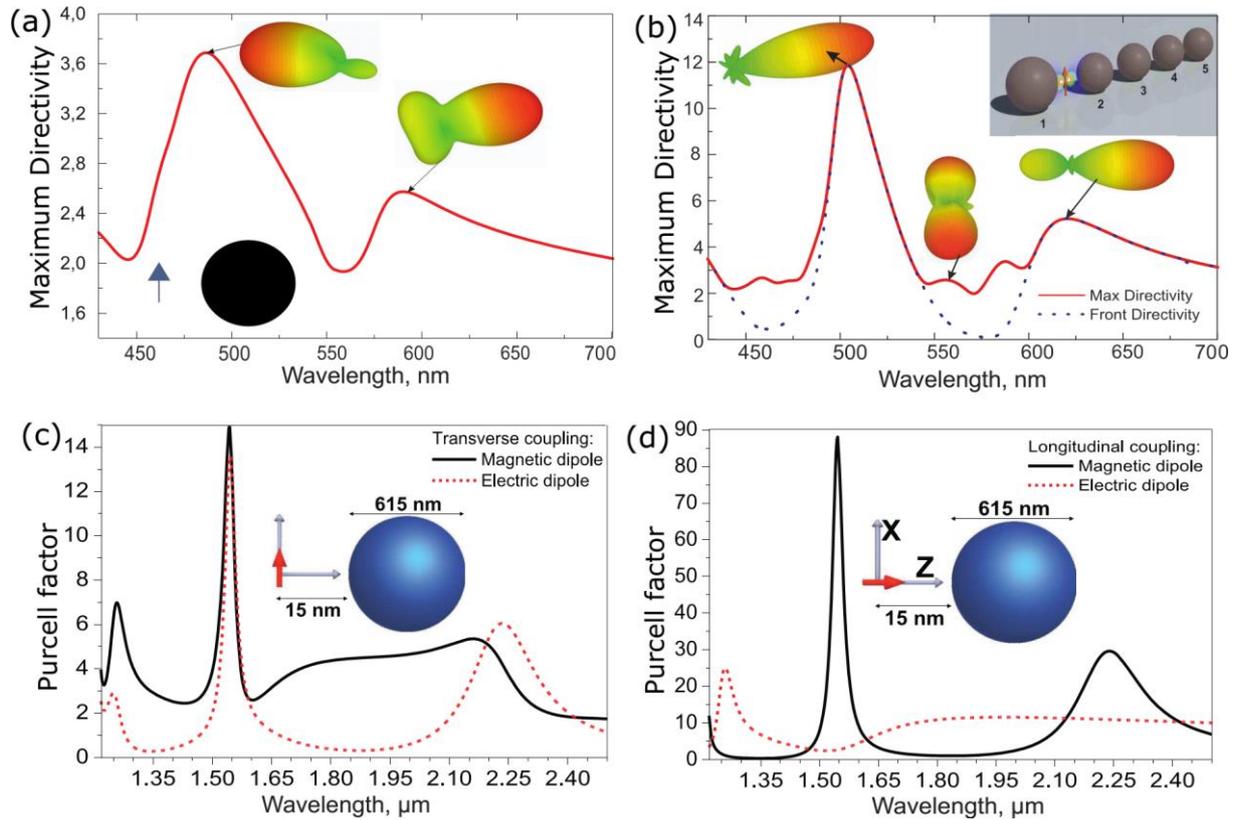

**Fugure 4.** Dielectric nanoantennas. (a) Directivity spectrum of the all-dielectric nanoantenna Huygens element, consisting of a Si NP and point dipole source. (b) Directivity spectrum of the all-dielectric Yagi-Uda nanoantenna, consisting of the reflector with the radius of 75 nm, and smaller directors with the radii of 70 nm. Insert: 3D radiation patterns at particular wavelengths. (c), (d) Normalized decay rates ($\gamma_{mol}/\gamma_{mol}^0$) spectra for an emitter placed at 15 nm from the surface of a Si sphere with diameter of 615 nm for transverse and longitudinal orientation of the dipole, respectively; the surrounding medium is air. Figures adapted with permission from: (a,b) -- Ref.[223], (c,d) -- Ref.[230].

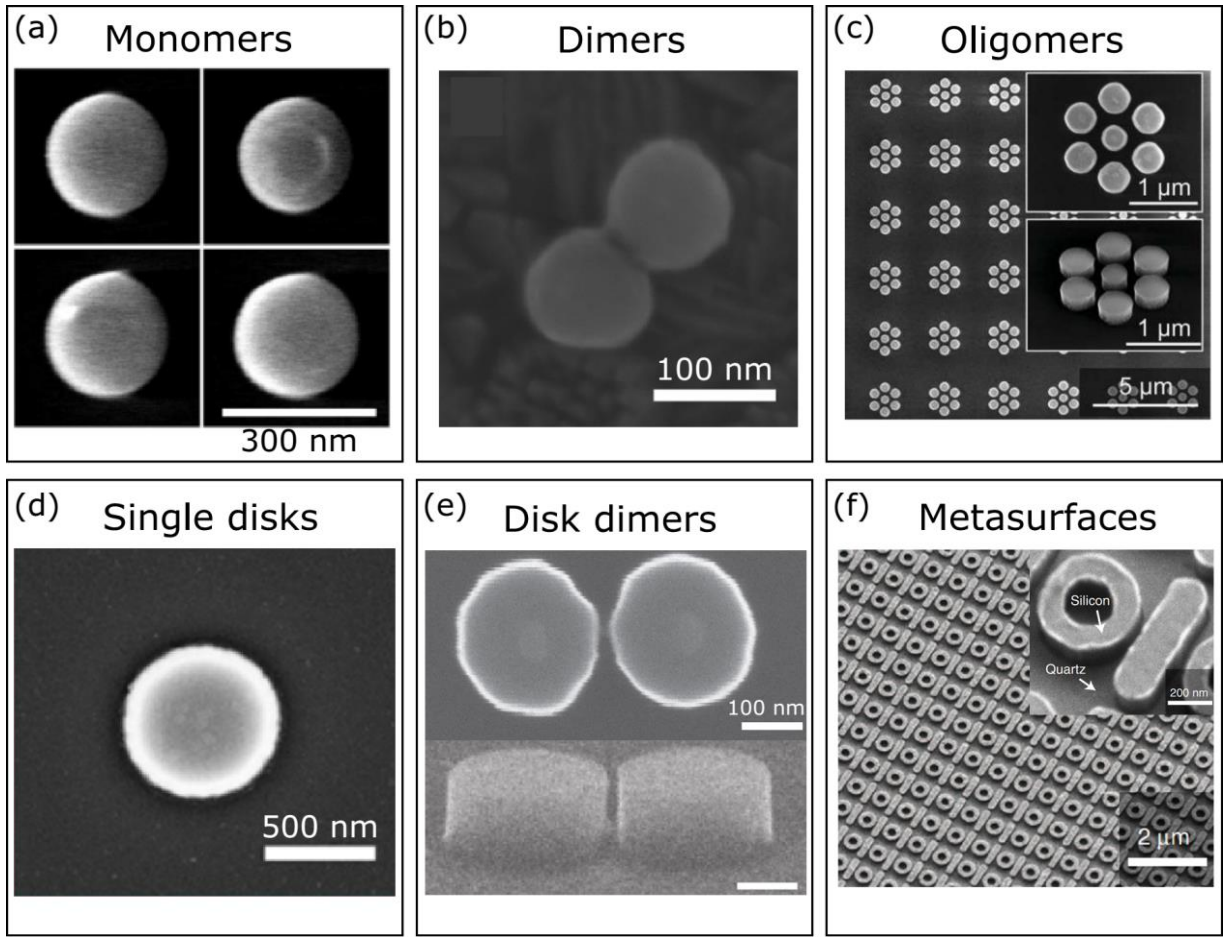

**Figure 5.** Examples of experimentally realized dielectric nanostructures. (a) Single spheres obtained by femtosecond-laser ablation. (b) Silicon sphere homodimers obtained by evaporation of a solution of silicon spheres obtained by femtosecond laser ablation. (c) Heptamer structure of silicon disks fabricated on a silicon oxide substrate. The surrounding disks have a constant diameter of 460 nm while the central disk is different for each oligomer. The inset shows a zoom on a single heptamer in top and oblique views. (d) Single Germanium disk on borosilicate glass using electron beam lithography. (e) Silicon disk dimers on a SOI substrate produced by electron beam lithography. Oblique view is also shown. (f) Example of an all-dielectric metasurface on a quartz substrate fabricated by electron beam lithography. Dimensions are shown in the figures with bars. Figures adapted with permission from: (a) -- Ref.[60], (b) -- Ref.[125], (c) -- Ref.[259], (d) -- Ref.[257], (e) -- Ref.[70], (f) -- Ref.[260].}

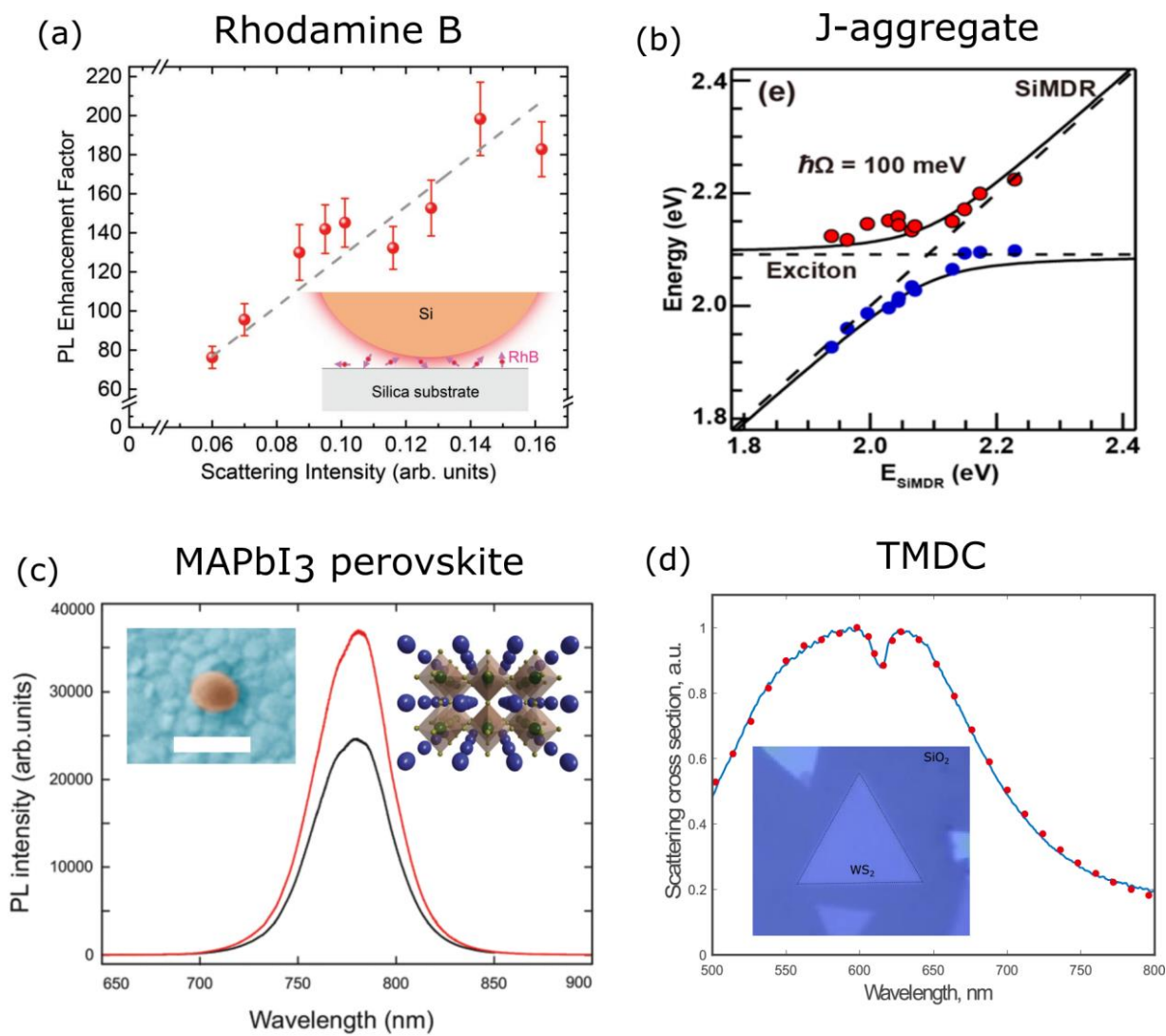

**Figure 6.** Different realized hybrid dielectric nanoparticle-excitonic systems. (a) PL enhancement factors in hybrid Si NP & *dye molecules* system as a function of corresponding scattering intensities obtained for nine single Si spheres. (b) Strong coupling regime (anticrossing) achieved in hybrid Si NP & *J-aggregate* system. The red and blue dots are extracted peaks from experimental dark-field scattering measurements. The black solid lines are fitting results according to the coupled harmonic oscillator model (see Eq. (11)). A mode splitting energy of 100 meV has been obtained. (c) PL intensity spectra from hybrid Si NP & *MAPbI$_3$ perovskite* system (red curve) and from reference MAPbI$_3$ perovskite (black curve). Left inset: colored SEM image of a perovskite film (blue area) with Si RDNs on its surface; the scalebar is 200 nm. Right inset: schematic of the three-dimensional interconnection of PbI$_6$-- octahedra in a perovskite lattice (green, Pb; yellow, I; blue, MA). (d) Scattering spectra in hybrid Si NP & monolayer WS$_2$ system. Inset: an optical image of monolayer WS$_2$. The values of coupling constant obtained from fitting of the experimental data demonstrate significant enhancement (from 24.8 meV up to 43.4 meV) with replacing air by water. Figures adapted with permission from: (a) - Refs.[90], (b) - Ref.[133], (c) - Ref.[86], (d) - Ref.[271].

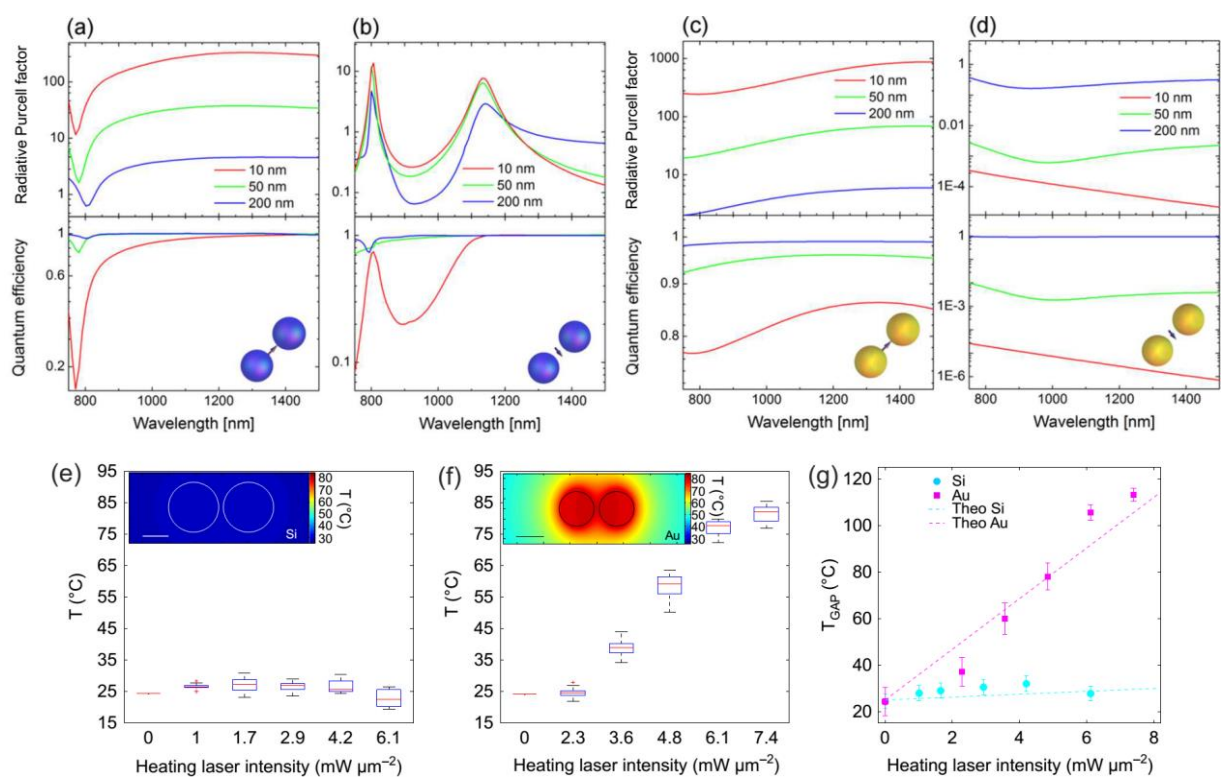

**Figure 7.** Comparison of the emission properties and temperature increase between Au and Si nanoantennas. Enhancement of the radiative decay rate and quantum efficiency of an electric dipolar emitter positioned in between two Si [(a), (b)] and Au [(c), (d)] nanospheres of 150 nm radius. Orientations of emitters positioned at the centers of the systems are shown in the schematics. Gap widths are given in the legends. (e), (f), (g) -- Temperature measurements over Au and Si nanoantennas. Box plot shows the average temperature T, measured for (e) Si and (f) Au nanoantennas, excited at the resonance. The inset in each figure shows the calculated temperature map around the disks for the heating laser intensity of 5 mW/μm$^2$ in both cases; scale bars are 100 nm. (g) Extracted temperature in the gap for selected silicon (cyan) and gold (magenta) nanoantennas as a function of the heating laser intensity at 860 nm. The dashed lines show the numerical calculations for the temperature at the gap, presenting good agreement with the experimental data. Error bars show the standard deviation of the temperature measurements, obtained from error propagation from the fluorescence measurements. Figure adapted with permission from Refs.[70,178].

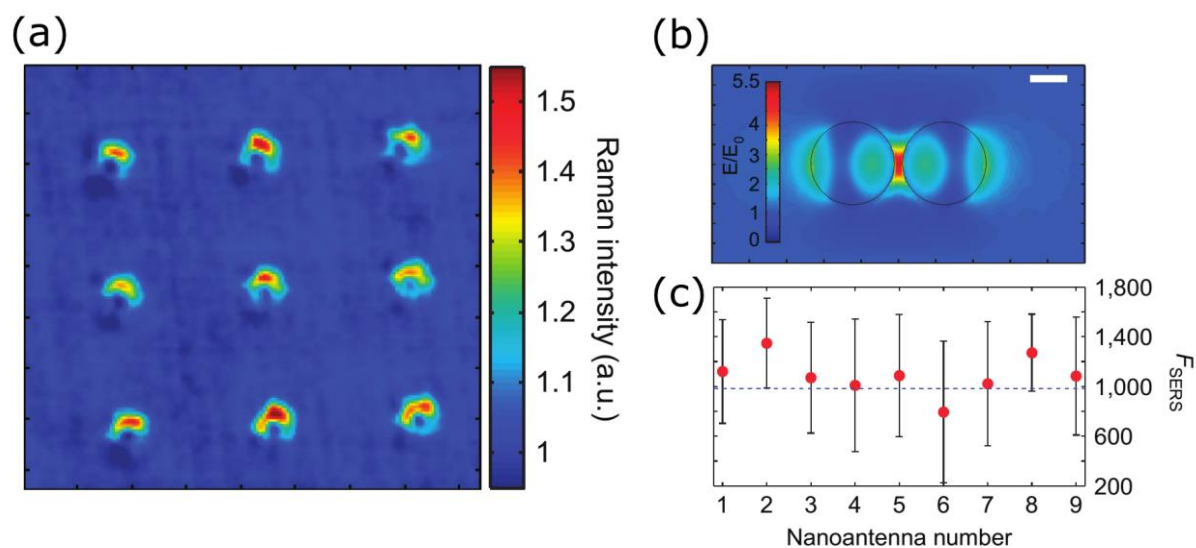

**Figure 8.** Surface enhanced Raman scattering (SERS) with all-dielectric dimer nanoantennas. (a) Experimental 2D normalized Raman map, showing the enhanced signal coming from molecules close to nanoantennas. (b) Near-field distribution map for the silicon structure excited at resonance, showing good confinement of the electric field in the 20nm-gap. (c) Experimental values for SERS obtained for each nanoantenna (dots) and the expected value (dashed line) based on the near-field intensity $(|\mathbf{E}|/|\mathbf{E}_0|)^4 \approx 1000$ from (b). The highest SERS enhancement factor reached in experiments is about 1400. Figure adapted with permission from Ref.[70].

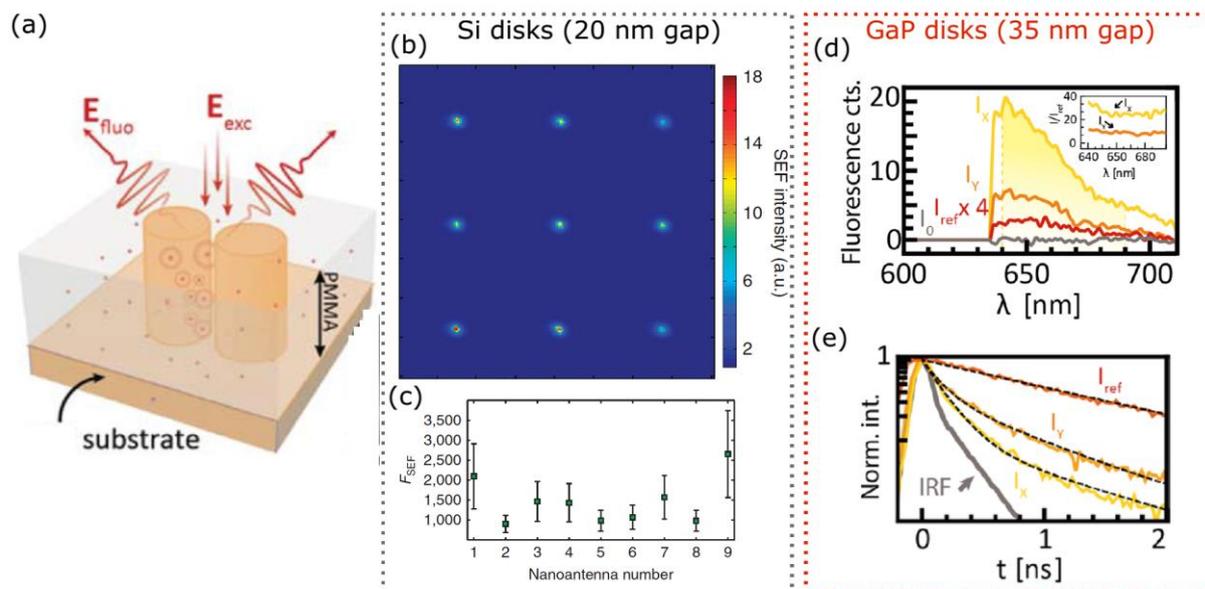

**Figure 9.** All-dielectric nanoantennas for many molecules fluorescence enhancement. (a) Experimental scheme: all-dielectric nanostructures are covered by a polymer matrix with dyes embedded. Many dye molecules cover the structure and some of them are at the hot-spot in the nanogap. Fluorescence excitation and detection is achieved in a epi-confocal configuration. (b) Experimental (normalized) fluorescence intensity map obtained for Si antennas with 20 nm gap. (c) Measured fluorescence enhancement factor obtained from the maximum values over each antenna in (b), accounting for the nanogap-enhanced volume. The error bars show half the difference between the minimum and the maximum value in each nanoantenna. (d) Comparison of the enhanced and initial fluorescence intensity spectra using GaP disks dimers with a gap of 35 nm. The subindex X and Y indicate the polarization of the exciting laser (X is parallel to the dimer). The reported enhancement factor is 3600 for parallel (X) and 150 for perpendicular (Y) excitation of the dimer. (e) Fluorescence lifetime decay curves, showing a lifetime reduction of at least 22x. Figures adapted with permission from Ref.[70] and Ref.[89].

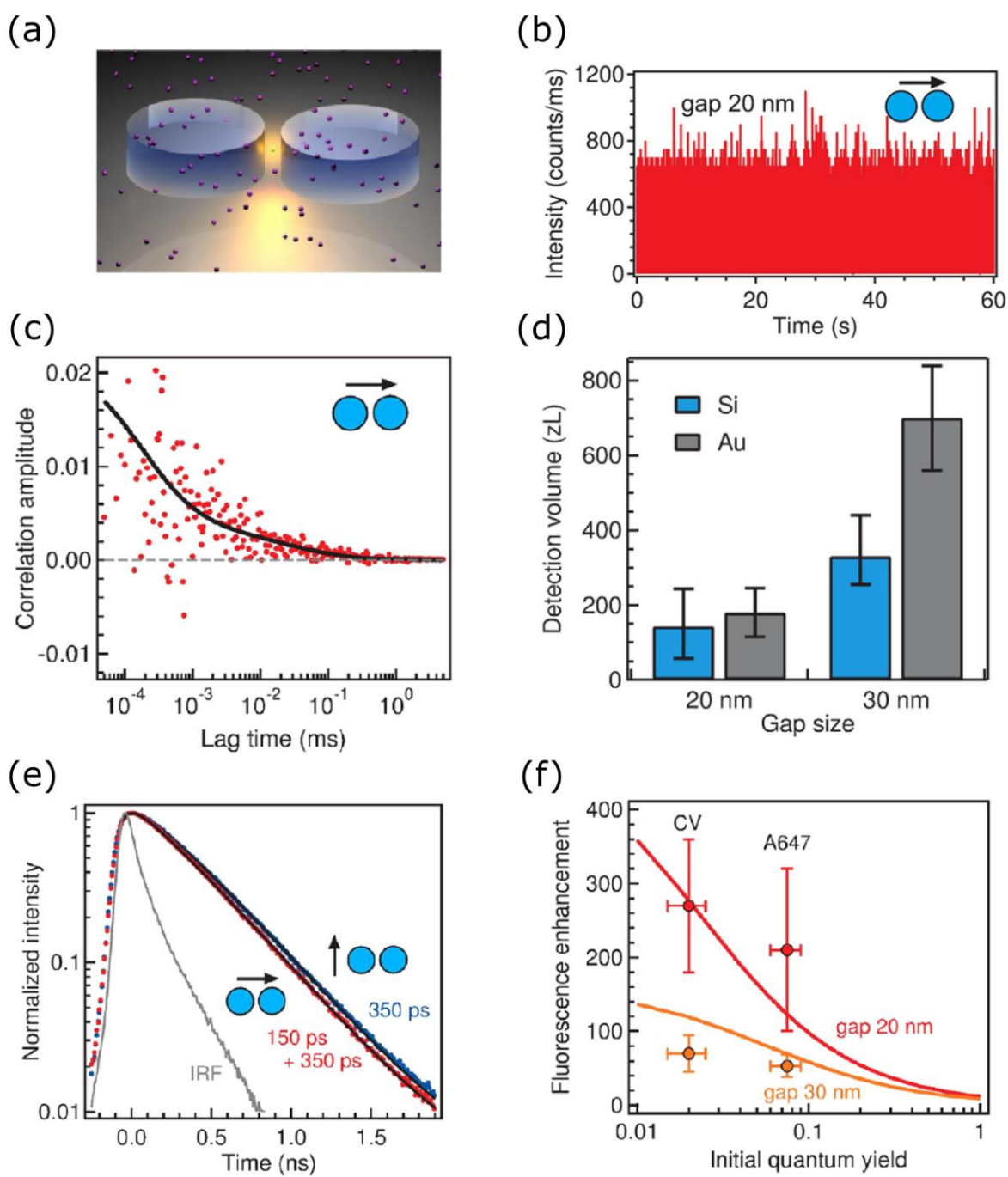

**Figure 10.** Single-molecule fluorescence enhancement with all dielectric nanoantennas. (a) Pictorial scheme of the experiment: fluorescent molecules diffuse around the silicon disks dimer while being excited. (b) Fluorescence (binned) time trace for 20 nm-gap dimer showing burst that correspond to SM fluorescence enhancement events. (c) Fluorescence correlation curve for the enhanced case with silicon 20 nm-gap. From the fit, the volume and the enhancement factor can be extracted. (d) Reduced detection volumes for silicon and gold

nanodimers extracted from the fluorescence correlation analysis. For the 20-nm gap, Si dimers show similar volumes as Au dimers while for 30 nm the silicon achieves a 2-fold reduction. (e) Fluorescence lifetime decay curve for parallel (red) and perpendicular (blue) to the dimer excitation polarization. (f) Overall fluorescence enhancement value for two dyes of different quantum yield for the two set of dimers with different gaps. These results show an excellent agreement with the theoretical calculations. Figure adapted with permission from Ref.[88].

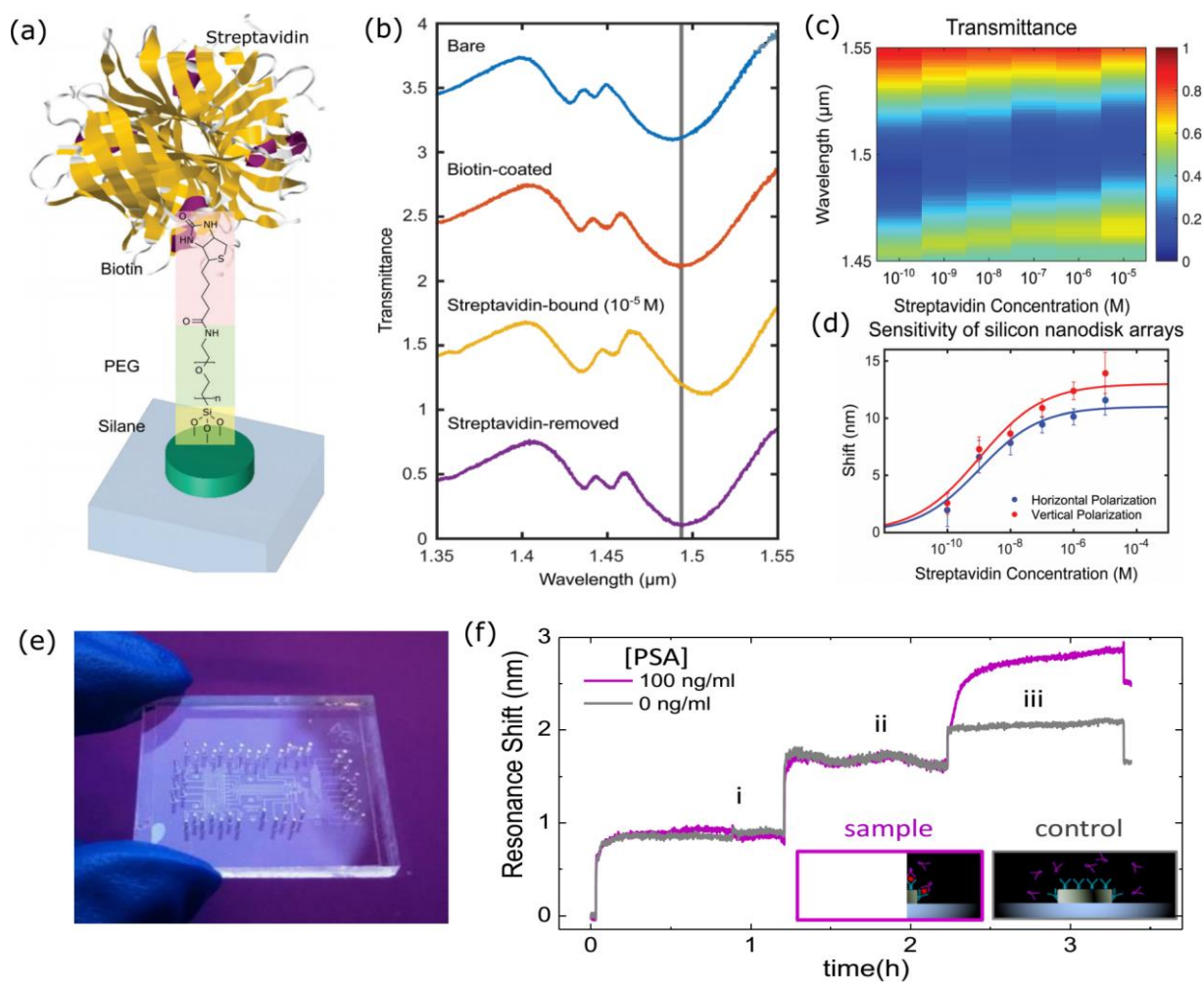

**Figure 11.** Biosensing applications with all-dielectric nanoantennas. (a) Operation principle of NP-based optical sensor on an example of streptavidin. (b) Optical transmittance spectra with vertically polarized light for the bare silicon nanodisks (blue), biotin-coated nanodisks

(orange) and streptavidin-bound nanodisks (yellow). (c) Color map of the transmittance spectra of nanodisk arrays measured with vertically polarized light for different streptavidin concentrations. (d) The average (mean) spectral shift of all nine arrays caused by the streptavidin (referenced to biotin-coated nanodisks) as a function of concentration. Polarization dependence is observed with vertical polarization (red dots) showing greater shifts than the horizontal polarization (blue dots). Error bars represent the standard deviation from all arrays. (e,f) On-a-chip biosensing with silicon nanodisks on quartz substrate. (e) Picture of an assembled chip with 8 sensing channels with silicon nanodisk arrays of different sizes. (f) Evolution of the nanodisk resonance during the different steps of the sandwich assay. The sample (100 ng/ml PSA) and control (no PSA) experiments are in grey and purple, respectively. Figure adapted with permission from: (a--d)--Ref.[157], (e, f)--Ref.[306].

| Material | Nanostructure | Type | Sensitivity, nm/RIU | FoM | Reference |
|---|---|---|---|---|---|
| Au | Sphere | ensemble | 90 | 1.5 | Underwood (1994)[318] |
| Au | Rod | ensemble | 239 | 1.8 | Paulo (2017)[319] |
| Au | Bipyramid | ensemble | 352 | 4.5 | Burgin (2008)[320] |
| Au | Rod-shape holes on film | metasurface | 588 | 3.8 | Liu 2010[321] |
| Au | Patch Nanohole | metasurface | 313 | 23.3 | Liu (2010)[305] |
| Si | Disks | ensemble | - | - | Bontempi (2017)[157] |
| Si | Disks | ensemble | 227 | 4.5 | Yavas (2017)[306] |
| Si | Double-gap split-ring | metasurface | 289 | 103 | Yang (2014)[260] |

**Table 1.** Comparison of the proposed spectroscopy and biosensor systems based on resonant dielectric and metallic nanostructures.

**Resonant dielectric nanoparticles made of materials with high dielectric permittivity become a powerful platform for modern light science, enabling various fascinating applications in nanophotonics and quantum optics**. Here, state-of-the-art applications of optically resonant high-index dielectric nanostructures for enhanced spectroscopies including fluorescence spectroscopy, surface-enhanced Raman scattering (SERS), biosensing, and lab-on-a-chip technology are surveyed.

**Keywords**: Resonant dielectric nanoparticles, nanophotonics, fluorescence spectroscopy, surface-enhanced Raman scattering, biosensing, lab-on-a-chip technology, Purcell effect, strong coupling, hybrid exciton-polariton systems

*Alex Krasnok\*, Martín Caldarola, Nicolas Bonod, and Andrea Alú\**

Dr. A. Krasnok, Prof. Dr. A. Alù
Department of Electrical and Computer Engineering, The University of Texas at Austin, Austin, Texas 78712, USA
E-mail: alu@mail.utexas.edu (A. A.), akrasnok@utexas.edu (A. K.)
Martín Caldarola
Huygens-Kamerlingh Onnes Laboratory, Leiden University, Leiden, Netherlands
Dr. Nicolas Bonod
Aix-Marseille Univ, CNRS, Centrale Marseille, Institut Fresnel, Marseille, France

**Spectroscopy and Biosensing with Optically Resonant Dielectric Nanostructures**

ToC figure:

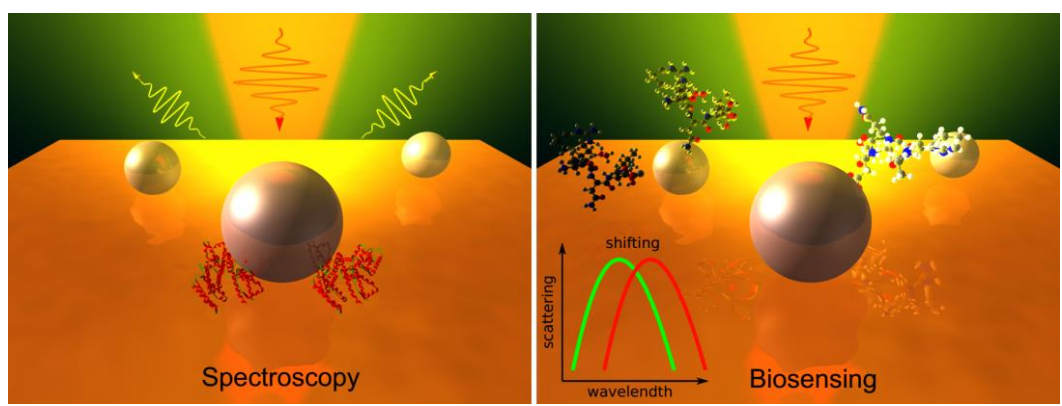